\documentclass{emulateapj}

\DeclareRobustCommand{\ion}[2]{%
\relax\ifmmode \ifx\testbx\f@series {\mathbf{#1\,\mathsc{#2}}}\else
{\mathrm{#1\,\mathsc{#2}}}\fi
\else\textup{#1\,{\mdseries\textsc{#2}}}%
\fi}

\slugcomment{To appear in ApJ.}

\shorttitle{PAH emission in nearby early-type galaxies}
\shortauthors{Vega et
al.}
\begin{document}
 \title{Unusual PAH emission in nearby early-type galaxies: \\
 a signature of an intermediate age stellar  population?}

  \author{O. Vega\altaffilmark{1}, A. Bressan\altaffilmark{1,2,3}, P. Panuzzo\altaffilmark{4}, R. Rampazzo\altaffilmark{2}, M. Clemens\altaffilmark{2}, G.~L. Granato\altaffilmark{5}, L. Buson\altaffilmark{2}, L. Silva\altaffilmark{5} and  W.~W. Zeilinger\altaffilmark{6}}

\altaffiltext{1}{Instituto Nacional de Astrof\'{\i}sica, Optica y
Electr\'onica, Apdos. Postales 51 y 216, C.P. 72000 Puebla, Pue.,
M\'exico.; ovega@inaoep.mx}

\altaffiltext{2}{INAF Osservatorio Astronomico di Padova, vicolo
dell'Osservatorio 5, 35122 Padova, Italy.}

\altaffiltext{3}{Scuola Internazionale Superiore de Studi Avanzati
(SISSA), via Beirut 4, 34014, Trieste, Italy.}

\altaffiltext{4}{CEA, Laboratoire AIM, Irfu/SAp, Orme des Merisiers, F-91191 Gif-sur-Yvette, France}
\altaffiltext{5}{INAF, Osservatorio Astronomico di Trieste, Via Tiepolo 11, I-34131 Trieste, Italy.}
\altaffiltext{6}{Institut fur Astronomie, Universitat Wien, Turkenschanzstrasse 17, A-1180
Wien, Austria.}


\begin{abstract}
{We present the analysis of \emph{Spitzer}--IRS spectra of four early-type galaxies, NGC~1297,
NGC~5044, NGC~6868, and NGC~7079, all classified as LINERs in
the optical bands. Their IRS spectra present the full series of H$_2$  rotational
emission lines in the range 5--38 $\mu$m, atomic lines, and prominent PAH features.
We investigate the nature and origin of the PAH emission, characterized by
 unusually low $6-9/11.3$ $\mu$m inter--band ratios. After the subtraction of a passive early type galaxy template, we find
that the $7-9$ $\mu$m spectral region  requires
dust features not normally present in star forming galaxies.
Each spectrum is then analyzed
with the aim of identifying their components and origin.}
{In contrast to normal star forming galaxies, where cationic PAH emission prevails, our 6--14 $\mu$m spectra seem to be dominated by large and neutral PAH emission, responsible for the low $6-9$/11.3 $\mu$m ratios,
plus two broad dust emission features peaking at 8.2 $\mu$m and 12 $\mu$m. Theses broad components, observed until now mainly in
evolved carbon stars and usually attributed to pristine material, contribute approximately 30--50\%
of the total PAH flux in the 6--14 $\mu$m region.}
{We propose that the PAH molecules in our ETGs arise from fresh carbonaceous material which is
continuously released by a population of carbon stars, formed in a rejuvenation
episode which occurred within the last few Gyr.  The analysis of the MIR spectra  allows us to infer
that, in order to maintain the peculiar size and charge distributions biased to large and neutral PAHs, this material
must be shocked, and excited by the weak UV interstellar radiation field of our ETGs.}


  \end{abstract}

 \maketitle
%

\section{Introduction}
\label{sec:introduction}
The understanding of the evolution of early-type  galaxies (i.e. E and
S0 galaxies)  has been greatly enhanced by the study of their
interstellar medium (ISM) \citep[see e.g.][for a review]{Renzini07}.
This component, and its relevance in early-type galaxies (ETGs
hereafter) was widely neglected in early studies, since these galaxies
were, for a long time, considered to be essentially devoided of
interstellar gas and dust.  In the last two decades, however,
multi-wavelength observations have changed this picture and have
detected the presence of a multi-phase ISM
\citep[e.g.][]{bregman92,Bettoni01,boselli05,Morganti06}. The bulk of
the gas in ellipticals is heated to the virial temperature, emitting
in X-rays \citep[e.g.][]{bregman92}.  Some ETGs also  contain  large
dust masses, which is surprising given the efficiency of sputtering by
their hot plasma \citep[e.g.][]{knapp89,ferrari99,temi04,2010arXiv1005.3056C}.
Even more surprising was the detection of polycyclic aromatic hydrocarbons
(PAHs) in ETG spectra
\citep[e.g.][]{xilouris04,kaneda05,kaneda08,bressan06,panuzzo07,bregman08},
because, dust molecules as small as PAHs, are expected to be more
easily destroyed by the interaction with the hot plasma
\citep[e.g.][]{dwek92}.

PAH emission is a typical feature of the mid-infrared (MIR hereafter) spectra
of star-forming galaxies
\citep[e.g.][]{roussel01,forster04,peeters04,vega05},  which are dominated by the strong  6.2, 7.7, and 8.6 $\mu$m emission bands usually  attributed to the C--C  stretching and C--H in-plane bending modes of ionized PAHs (see Tielens 2008 and references therein).  However, the  PAH spectra of ETGs are
often different from those of star forming galaxies.
The most striking characteristic is the unusual PAH
inter--band emission ratio between the 7.7 $\mu$m complex
and the 11.3 $\mu$m complex \citep{kaneda05,bressan06}. So,
while in star forming galaxies the 6.2, 7.7, and 8.6 $\mu$m features dominate
the PAH spectrum, they are very weak in  these ETGs,
in spite of a strong 11.3 $\mu$m band.

There already exist studies in the literature aimed at understanding
the nature and origin of these PAH emission features, as well as
the physical conditions of the ISM,
that allow the survival of this material in the observed ratios. \citet{smith07}
 analyzing the SINGS galaxies sample,
found that some LINERs also have unusual PAH ratios. They suggest that
these ratios could be due to the relatively hard radiation field from a
low-luminosity AGN, which modifies the emitting grain distribution
at the expense of the smaller PAHs.
\citet{bregman08} claim that, once the continuum of the underlying stellar population, which has a
broad dip just below the
7.7$\mu$m complex, is properly
subtracted, the low 7.7 to 11.3 $\mu$m ratio turns into a value typical of the ISM of
late-type galaxies.
In contrast, \citet{kaneda08} have shown cases where
these ratios remain anomalous even after the proper removal of the
underlying stellar population.
These authors conclude that the anomalous ratios are likely due to the presence of
a larger fraction of neutral PAHs  with respect to
ionized PAHs, contrary to what happens in the ISM of late type galaxies.
They ascribe this anomalous mixture to recent accretion events.

In this paper we deepen the investigation of this problem by
presenting a new approach to the study of these peculiar ETGs.  We
analyze our, high S/N \emph{Spitzer}--IRS  observations of four
ETGs (see Table \ref{tableape} for the galaxy classification) in low
density environments (LDEs), namely  NGC~1297, NGC~5044, NGC~6868, and
NGC~7079, which display ``unusual" PAH emission in their MIR
spectra. For our study we make use of  theoretical spectra and
empirical dust feature templates recently  reported in the literature
(see e.g. Rapacioli et al. 2005, Bern\'e et al. 2007, Joblin et
al. 2008; Bauschlicher et al. 2008, Joblin et al. 2009, Bern\'e et
al. 2009).  This method has the advantage of isolating the components
responsible for the emission, thus providing more insight into the origin
and evolution of this ISM component \citep{allamandola99}.  We
advance a new scenario, where this PAH emission is naturally explained
in terms of the presence of an underlying intermediate age population
of carbon stars, which are continuously feeding the ISM with pristine
carbonaceous material.  If our interpretation is correct, the
anomalous PAH spectra are strong  evidence of a rejuvenation episode
that affected the evolution of the ETGs in the recent past.

The paper is organized as follows. In \S~2 we provide the details of
the observations and data reduction methods. In \S~3 we measure and
provide detailed information for each single PAH emission feature
observed in the spectra and for PAH inter-band ratios. This
information is used to compare our spectra with existing
classifications.  In \S~4 we explore the nature of the PAH carriers by
using component templates available both from observations and
theory. This method allow us to identify families of emitters typical
of different environments. In \S~5 we analyze the MIR emission
  lines sensitive to the physical conditions of the medium giving
  place to the PAH emission.  In \S~6 we discuss our results and
compare them with other studies.  The relevant properties of our ETGs,
as known so far, are provide in the Appendix.

\section{Mid Infrared Observations and data reduction}
\label{sec:reduction}
\emph{Spitzer} IRS spectral observations of all galaxies were obtained
during the third \emph{ Spitzer} General Observer Cycle~3 on 2007 June 1 as
part of program [ID 30256;  P.I.: R.Rampazzo].

The observations were performed in Standard Staring mode with low
resolution ($R\sim$ 64--128) modules SL1 (7.4--14.5$\mu$m), SL2
(5--8.7$\mu$m), LL2 (14.1--21.3$\mu$m), and LL1 (19.5--38$\mu$m).
Exposure times for each galaxy are given in Table~\ref{table2}.

We have devised an {\it ad hoc} procedure to
flux calibrate the spectra that exploits the large degree of symmetry
that characterizes the light distribution in early-type galaxies. The
reduction procedure is fully described in \citet{bressan06}; here
we only summarize the main steps.

For SL observations, the sky background was removed by subtracting
observations taken in different orders, but at the same nod position.
LL segments were sky-subtracted by differencing the two nod positions.

We obtained new e$^-$~s$^{-1}$ to Jy flux conversions by applying a correction
for aperture losses (ALCF) and a correction for slit losses (SLCF) to
the flux conversion tables provided by the \emph{Spitzer} Science Center
\citep[e.g.][]{kenn03}. By applying the ALCF and SLCF corrections we
obtained the flux received by the slit.

\begin{table}
\centering
\caption{IRS observations}
\begin{tabular}{lccc}
\hline
 Galaxy       & ramp duration  &  number of cycles&AOR Key \\
  IRS module     &       [seconds]                &                    \\
\hline
& & \\
NGC 1297          &   & &17923584\\
& &  \\
SL1   & 60 & 19 \\
SL2      &60 &  19\\
LL2    &120 & 14 \\
LL1      &120 & 8 \\
& & \\
NGC 5044          &   & &17926144\\
& &  \\
SL1   & 60 & 19 \\
SL2      &60 &  19\\
LL2    &120 & 14 \\
LL1      &120 & 8 \\

& & \\
NGC 6868          &  & &17926912\\
& &  \\
SL1   & 60 & 6 \\
SL2      &60 &  6\\
LL2    &120 & 13 \\
LL1      &120 & 8 \\
& & \\
NGC 7079          &   &&17927168 \\
& &  \\
SL1   & 60 & 19 \\
SL2      &60 &  19\\
LL2    &120 & 14 \\
LL1      &120 & 8 \\
\hline
\end{tabular}
\label{table2}
\end{table}

For each galaxy we then simulated the corresponding observed
profile along the slits by convolving a wavelength dependent
bidimensional intrinsic surface brightness profile with the instrumental
point spread function (PSF). The adopted profile was a two dimensional
modified King law \citep{elso87}. By fitting the observed profiles with
the simulated ones, we could reconstruct the intrinsic profiles and the
corresponding intrinsic spectral energy distribution (SED).
This procedure allowed us to determine whether a particular feature was
spatially extended or not.

Finally the spectrum was extracted in a fixed width (18\arcsec\ for SL
and 10.2\arcsec\ for LL) around the maximum intensity. After normalization
to the same area (3.6\arcsec$\times$18\arcsec) there remain differences between
the SL and LL spectral segments of between 3$\%$ (NGC~5044) and 14$\%$ (NGC~6868).
In order to account for these differences, the LL spectrum has been
rescaled to match the SL one in the common wavelengths.

The uncertainty on the flux was evaluated by considering two sources of
noise: the instrumental plus background noise and the Poissonian noise
of the source. The former was evaluated by measuring the variance of
pixel values in background-subtracted, co-added images far from the
source. The Poissonian noise of the sources was estimated as the square
root of the ratio between the variance of the number of e$^-$ extracted
per pixel in each exposure, and the number of the exposures. The total
noise was obtained by summing the two sources in quadrature and by
multiplying by the square root of the extraction width in pixels. We
notice that the overall absolute photometric uncertainty of IRS is 10 per cent,
while the slope deviation within a single segment (affecting all spectra
in the same way) is less than 3 per cent (see the \emph{Spitzer} Observers Manual).

\begin{figure}
\includegraphics[width=0.5\textwidth]{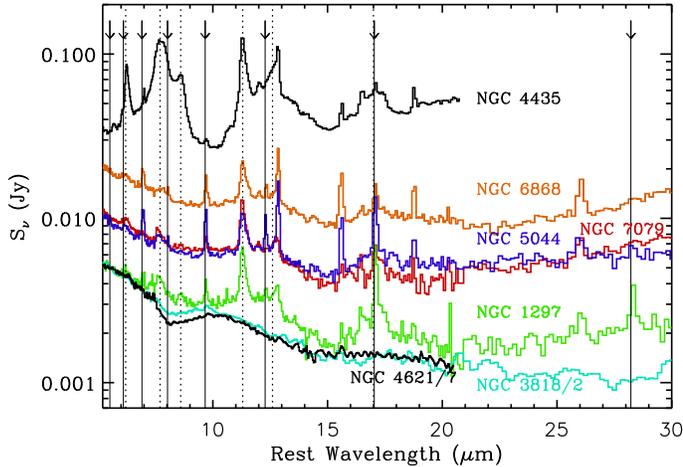} \caption{\emph{Spitzer}-IRS spectra of NGC~1297, NGC~5044, NGC~6868,
and NGC~7079. Dotted lines indicate the positions of the most
relevant PAH complexes at 6.2 $\mu$m, 7.7 $\mu$m, 8.6 $\mu$m, 11.3
$\mu$m, 12.7 $\mu$m, and 17 $\mu$m.  Arrows and thin solid lines
indicate  the positions of the H$_2$ rotational lines. For comparison, we  also show the spectra of an ETG with a recent
burst of star formation (NGC~4435, Panuzzo et al. 2007) and of two
passively evolving ETGs: the cluster ETG, NGC~4621 (Bressan et al. 2006a), and the field ETG, NGC~3818 \citep{Panuzzo10}, both scaled to the flux of
NGC~1297 at $5.6 \mu$m.} \label{fig1}
\end{figure}

\begin{figure*}
\includegraphics[width=0.5\textwidth]{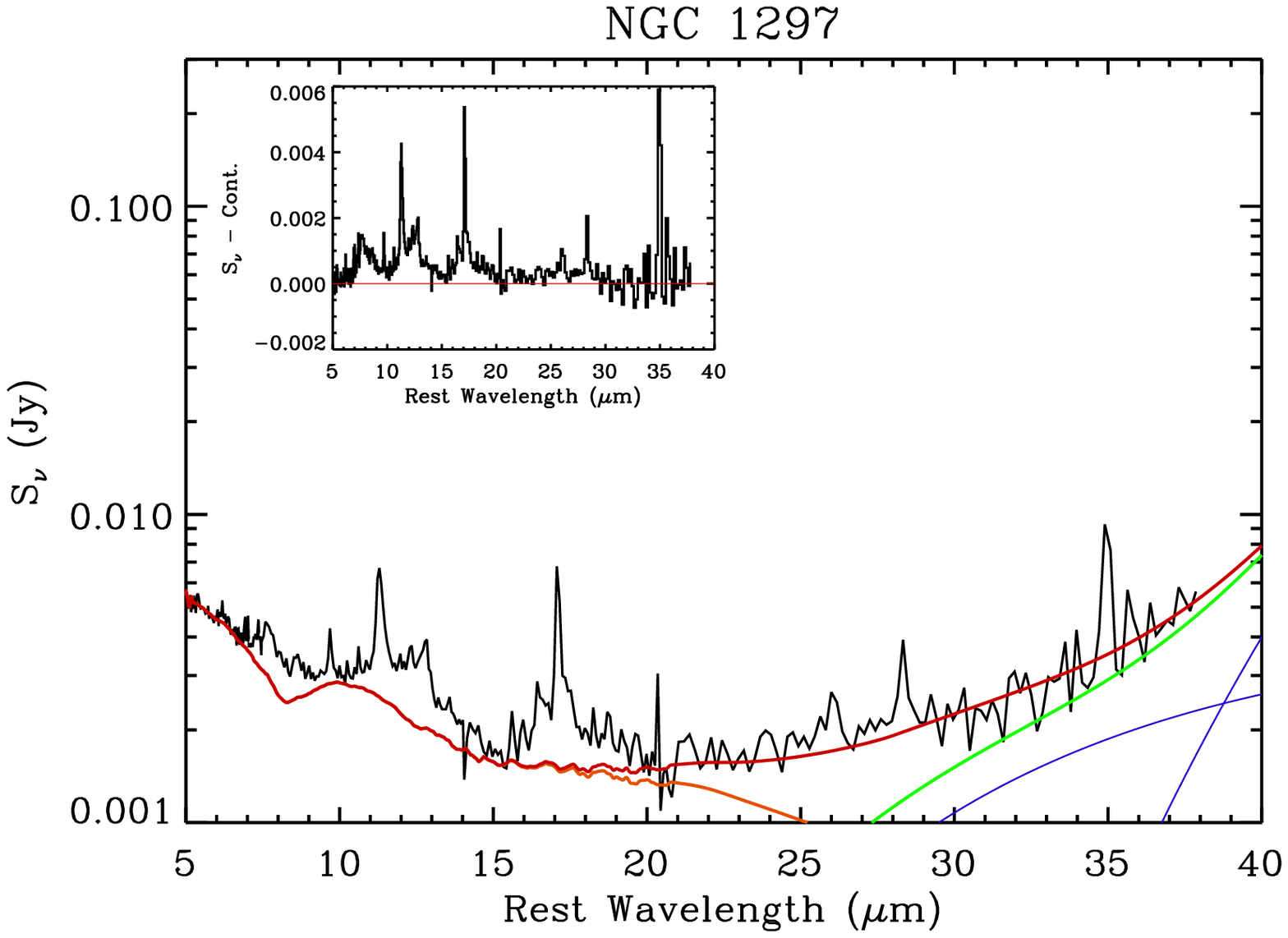}
\includegraphics[width=0.5\textwidth]{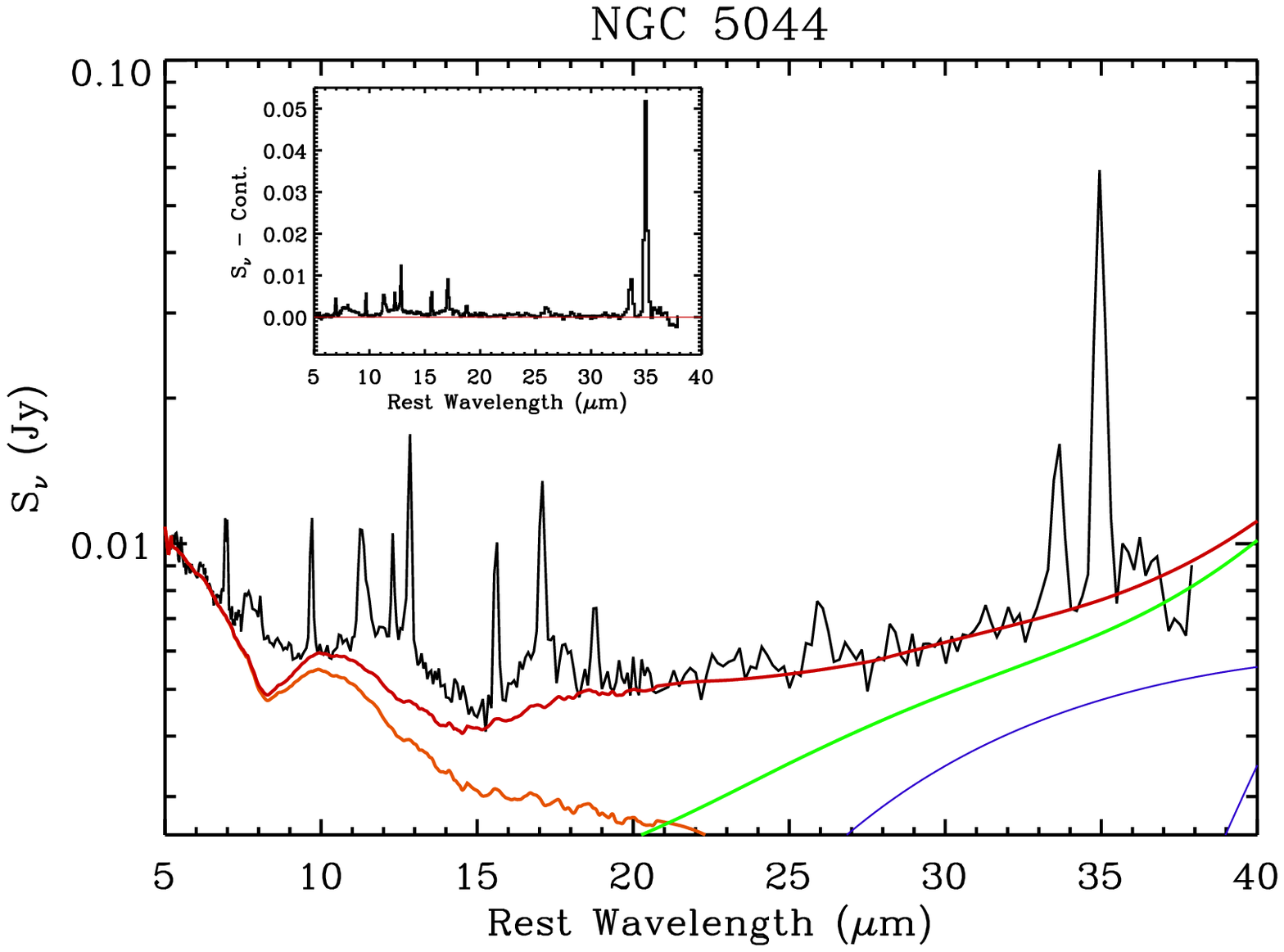}
\includegraphics[width=0.5\textwidth]{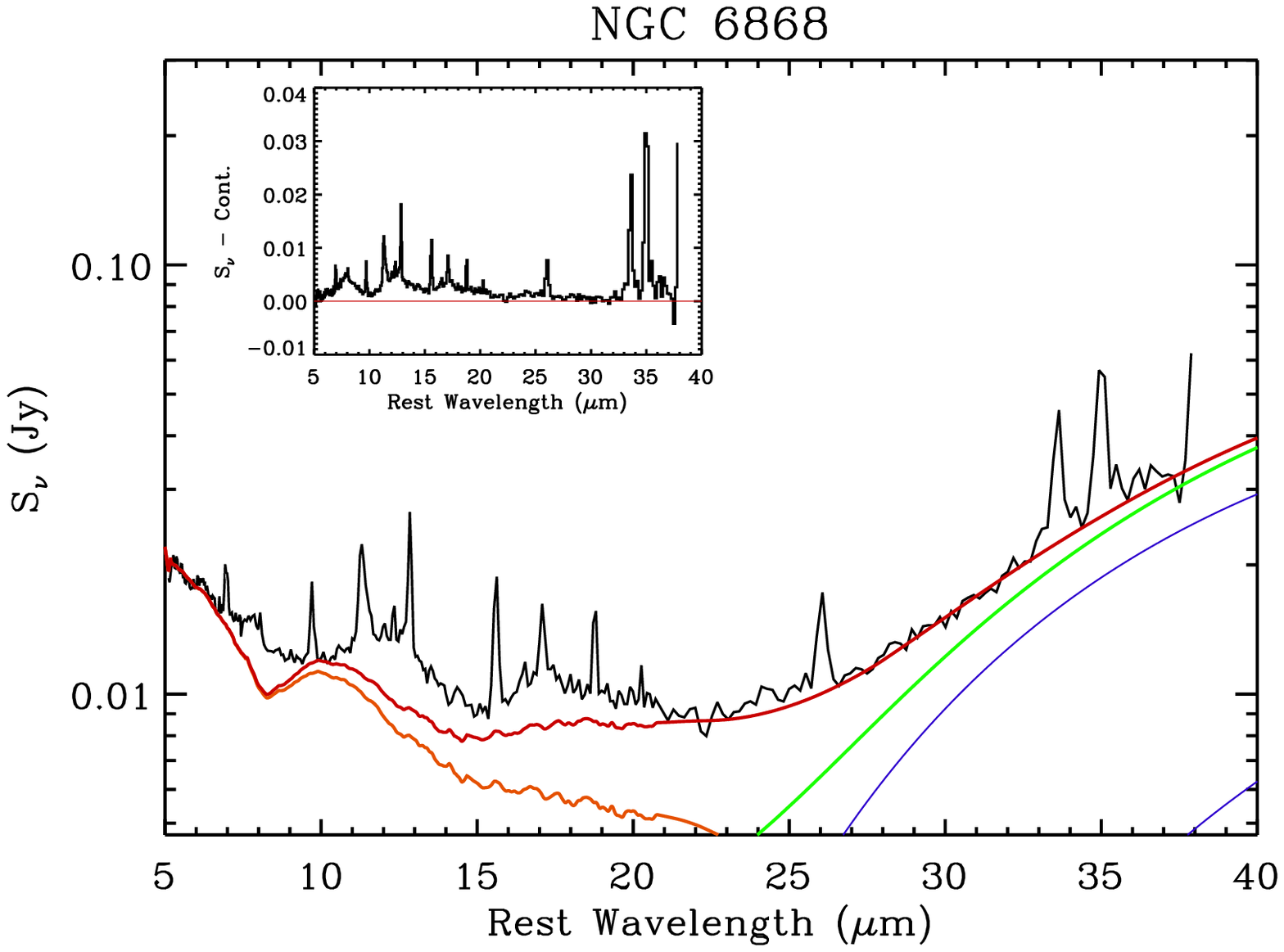}
\includegraphics[width=0.5\textwidth]{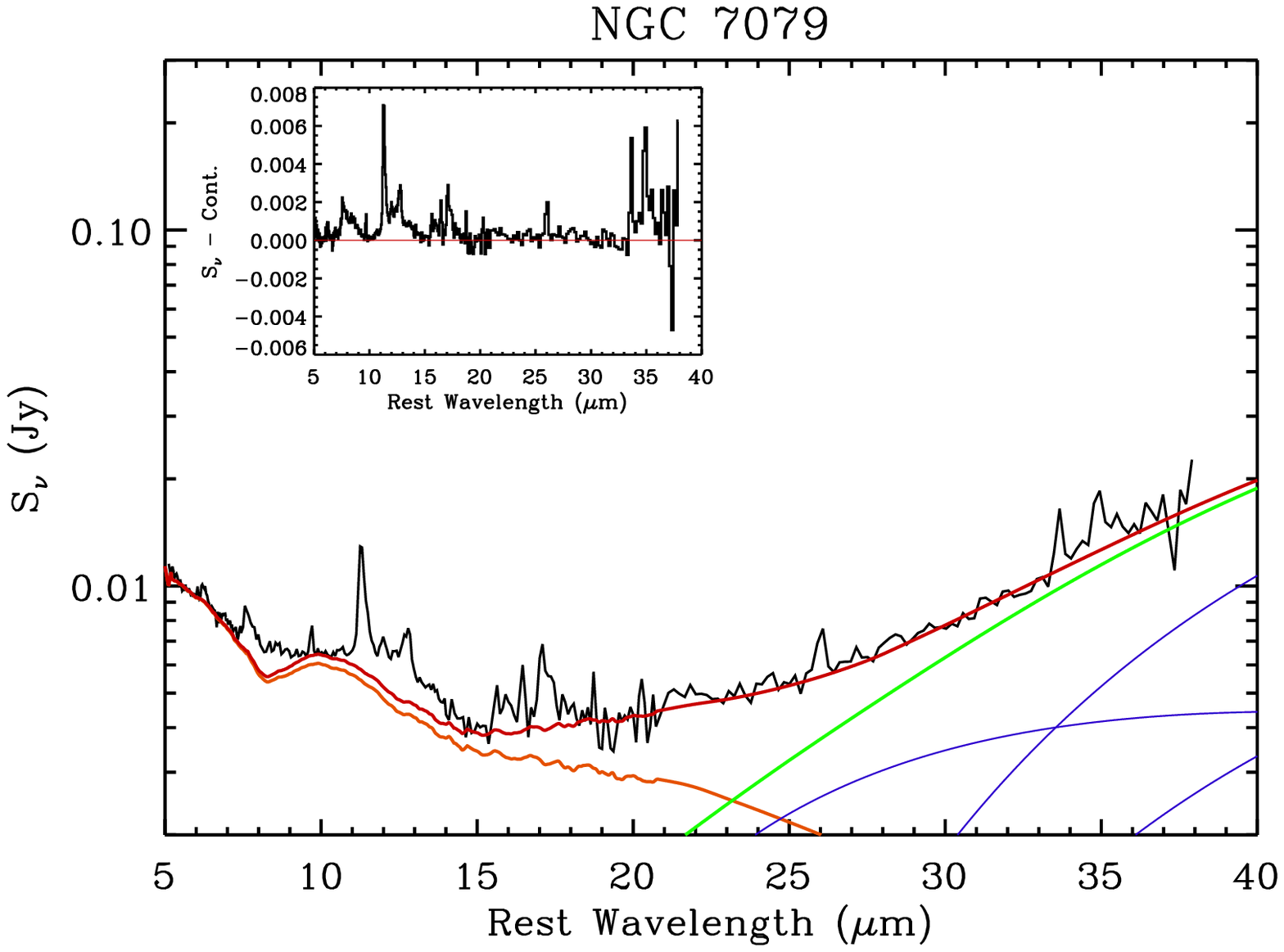}
\caption{Detailed fit to the MIR continuum emission of our sample of
ETGs. The orange line represents the template of the passive
ETG normalized to the H-band. Blue lines represent the  different thermal dust
components. The
green line represents the sum of all  dust components. The red line is the best fit model to the total
continuum emission, calculated as the  sum of the normalized passive
ETG plus the thermal dust components. In the insets we plot the
continuum subtracted spectra. Notice the broad feature around 8
$\mu$m that emerges when the continuum is subtracted.} \label{fig2}
\end{figure*}

\begin{figure*}
\includegraphics[width=0.9\textwidth]{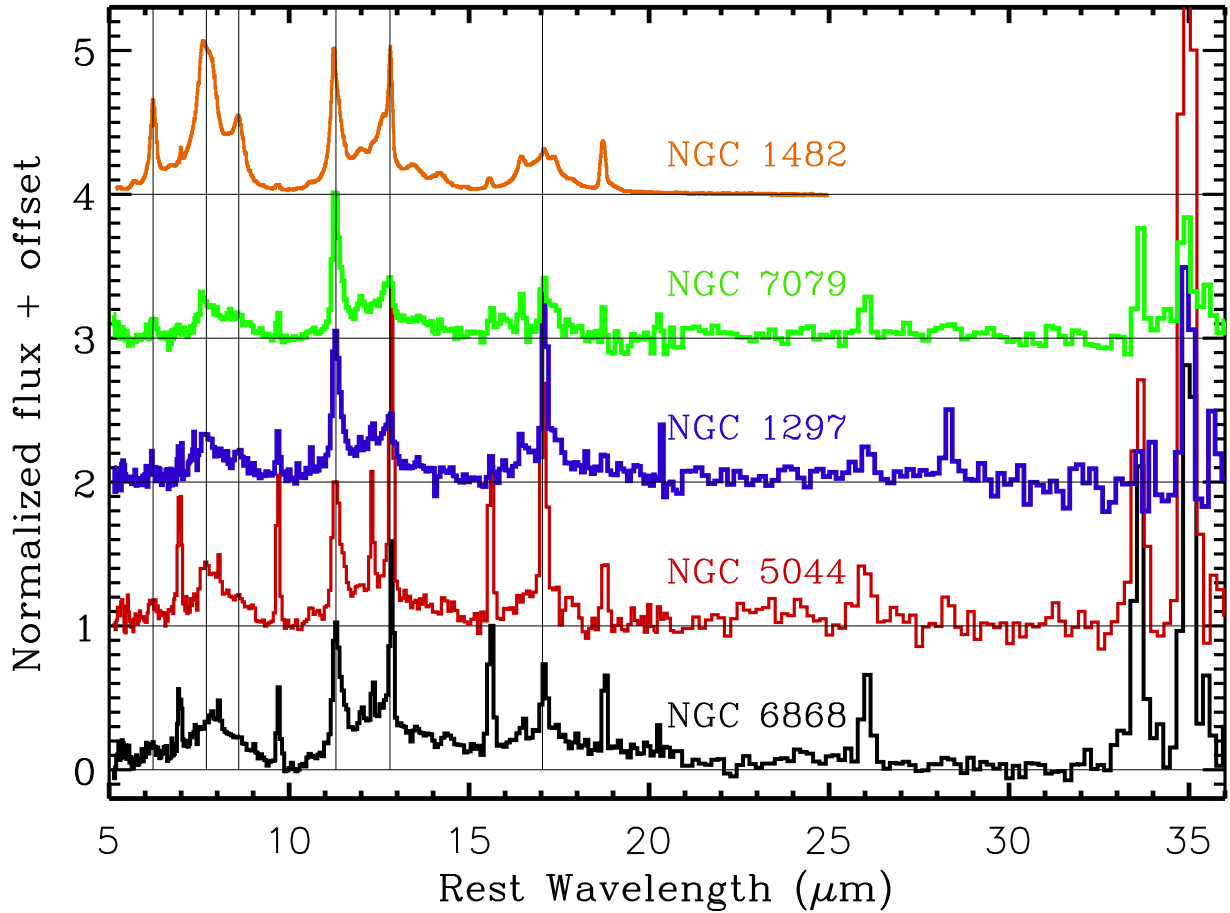}
\caption{Continuum-subtracted spectra of our ETGs  normalized to the peak intensity of the 11.3 $\mu$m feature. For comparison, we also show the IRS spectrum of NGC~1482, a SINGs HII galaxy from \citet{smith07}. Vertical lines indicate the positions of the main PAH complexes at 6.2, 7.7, 8.6, 11.3 12.8, and 17 $\mu$m.  } \label{fig5}
\end{figure*}

\begin{figure*}
\includegraphics[width=0.5\textwidth]{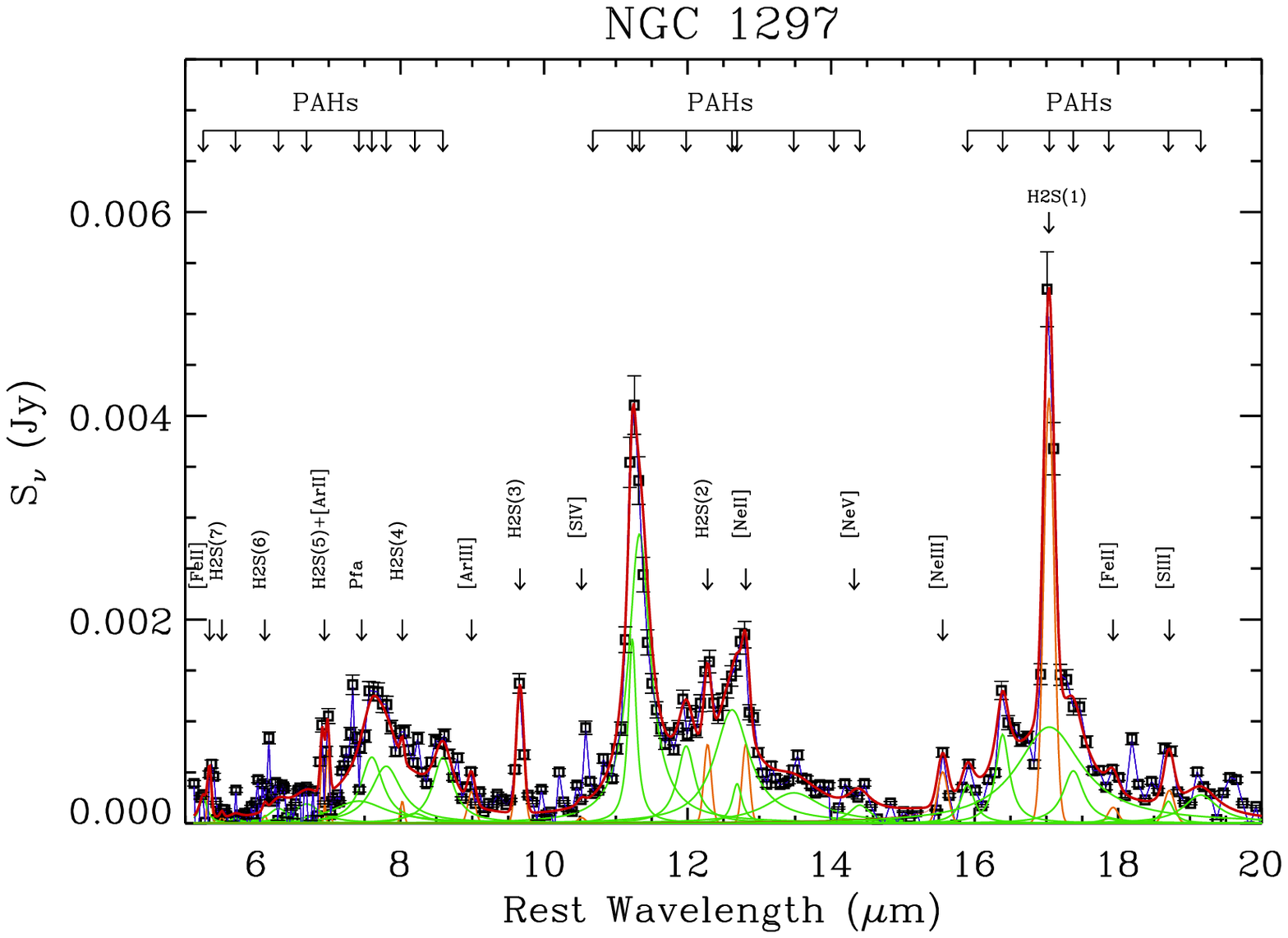}
\includegraphics[width=0.5\textwidth]{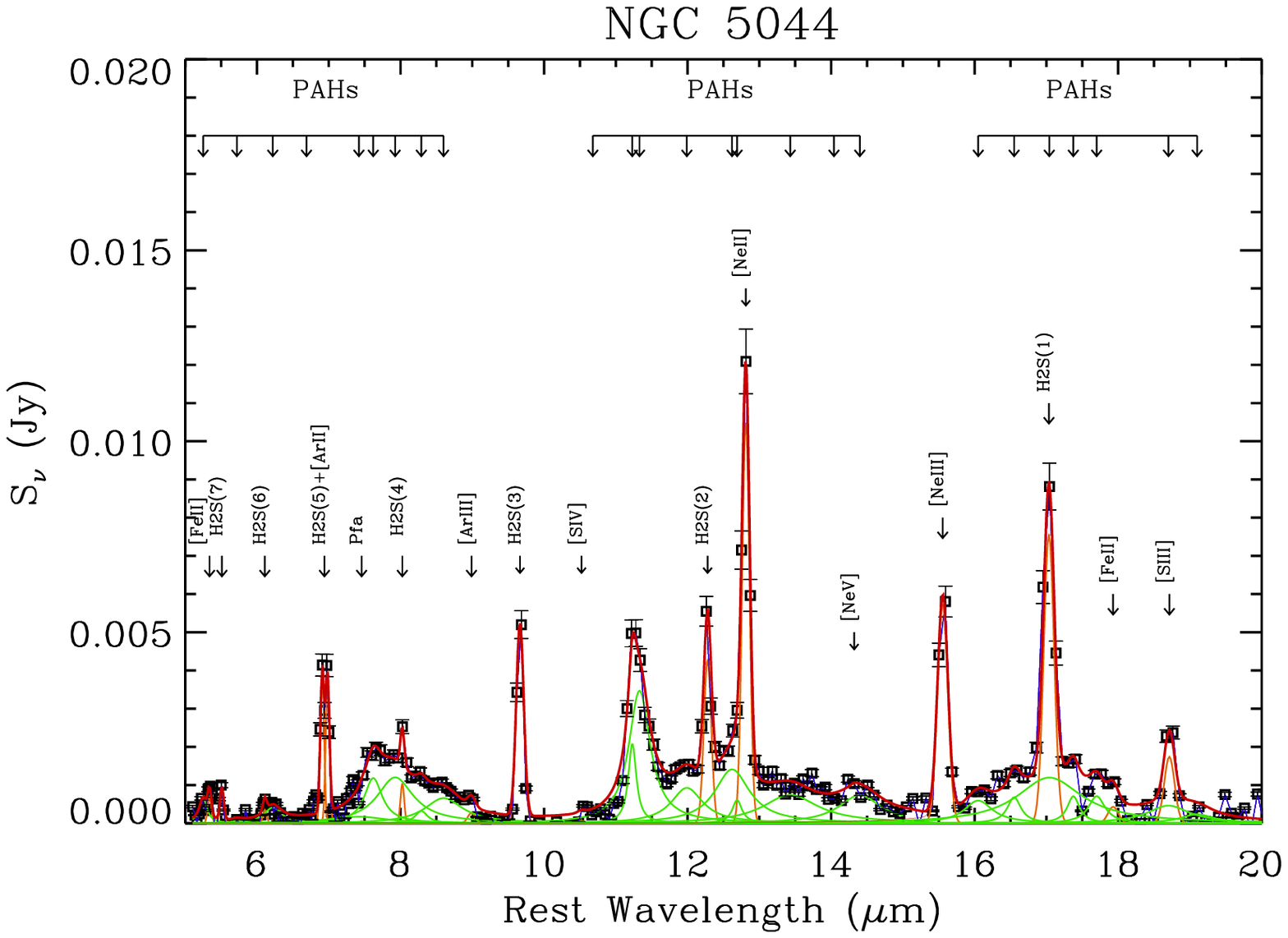}
\includegraphics[width=0.5\textwidth]{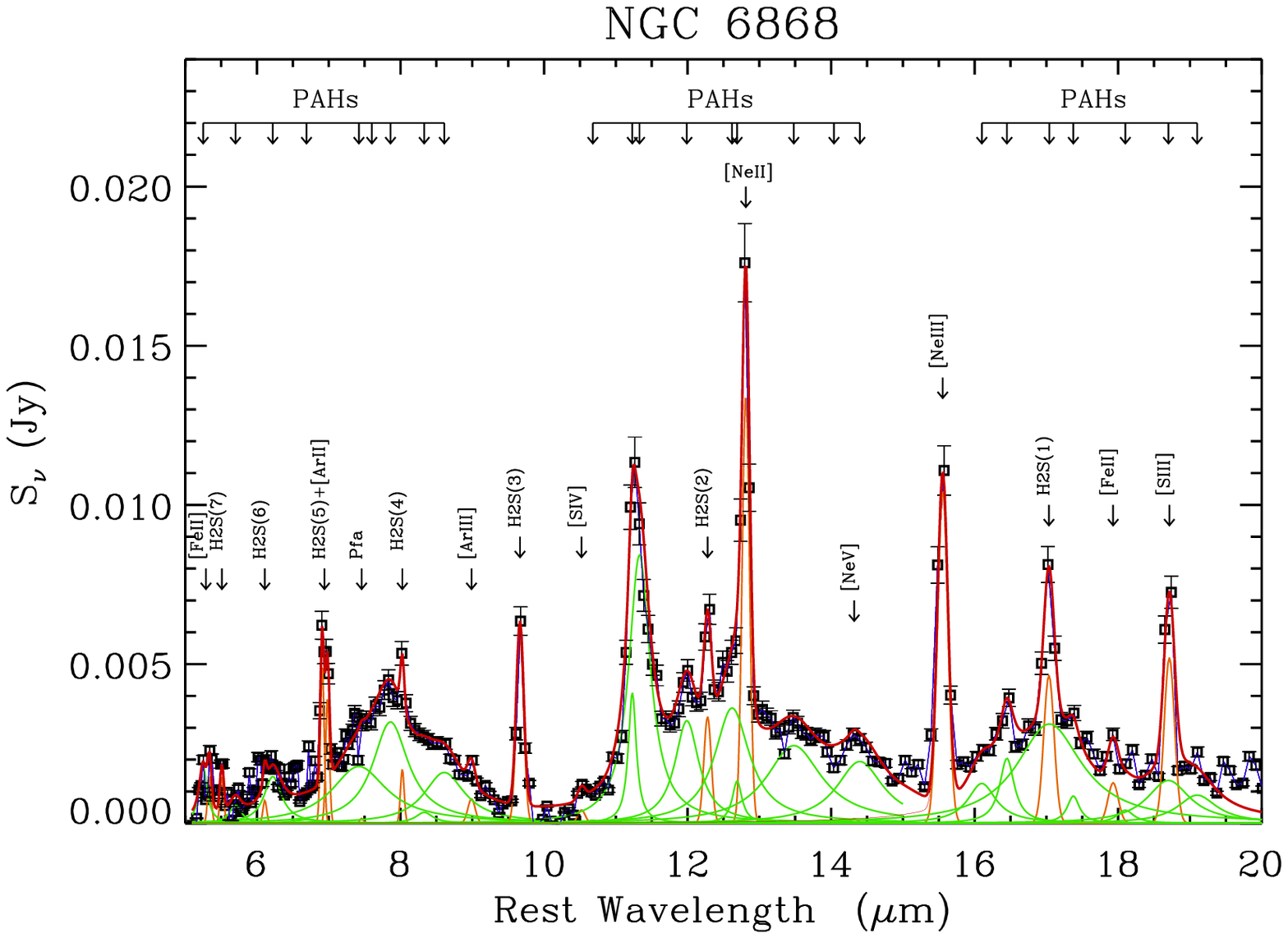}
\includegraphics[width=0.5\textwidth]{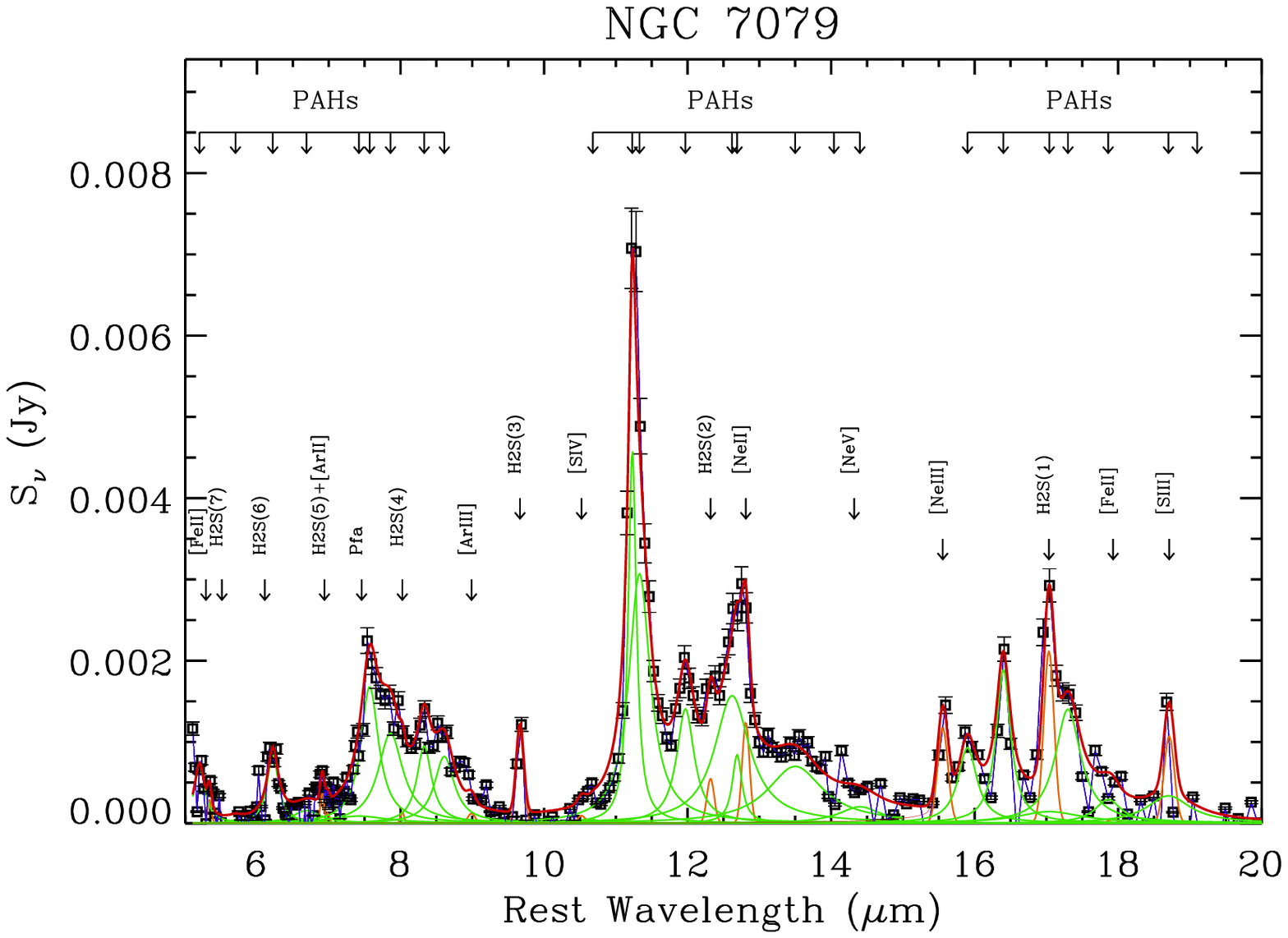}
\caption{Detailed fit to the lines and the PAH emission  for
the continuum subtracted spectra of NGC~1297 (upper left), NGC~5044
(upper right), NGC~6868 (lower left), and NGC~7079 (lower
right) in the spectral region between 5 and 20 $\mu$m.  The  open squares
and the blue solid line indicate the observed MIR continuum
subtracted spectra. The solid red line indicates the total fit to the
data, calculated as the sum of the PAH features (solid thin green lines), and
the emission lines (solid thin orange lines). Arrows indicate the
positions of the fitted PAH features, nebular emission lines, and H$_2$
rotational lines.} \label{fig3}
\end{figure*}

\begin{table}
\centering
\caption{2MASS magnitudes within the central 5 arcsec radius}
\begin{tabular}{lccc}
\hline
NAME&J&H&K\\
\hline
NGC~1297&11.98$\pm0.01$&11.24$\pm0.01$&11.01$\pm0.01$\\
NGC~5044&11.24$\pm0.01$&10.51$\pm0.01$&10.22$\pm0.01$\\
NGC~6868&10.41$\pm0.01$&9.67$\pm0.01$&9.40$\pm0.01$\\
NGC~7079&11.10$\pm0.01$&10.41$\pm0.01$&10.17$\pm0.01$\\

\hline
\end{tabular}
 \label{tablenir}
\end{table}


\begin{turnpage}
\begin{table*}
\centering
\caption{Parameters and integrated fluxes of the fitted PAH
features.}
\begin{tabular}{lc|lcr|lcr|lcr|lcr}
\hline\hline
{DL07}$^{*}$& {}&{}&{N1297}$^{\natural}$&{}&{}&{N5044}$^{\natural}$&{}&{}&{N6868}$^{\natural}$&{}&{}&{N7079}$^{\natural}${}\\
{$\lambda_{\rm 0}^{(e)}$}&{$\gamma$}& {$\lambda_{\rm 0}^{(e)}$} &
{$\gamma$} & {F$^{(f)}$}& {$\lambda_{\rm 0}^{(e)}$} & {$\gamma$} &
{F$^{(f)}$}& {$\lambda_{\rm 0}^{e}$} & {$\gamma$} & {F$^{(f)}$}&
{$\lambda_{\rm 0}^{(e)}$} &
{$\gamma$} & {F$^{(f)}$} \\
\hline
5.27&0.034&5.25$\pm0.01$&0.025$\pm0.006$&5.76$\pm2.71$&5.25$\pm0.01$&0.025$\pm0.002$&13.38$\pm5.37$&5.25$\pm0.01$&0.025$\pm0.005$&40.51$\pm15.04$&  5.20$\pm0.01$&0.034$\pm0.006$&22.13$\pm10.17$ \\
5.70&0.035&5.70$\pm0.01$&0.035$\pm0.015$&1.37$\pm4.45$&5.72$\pm0.01$&0.035$\pm0.015$&8.41$\pm5.17$&5.70$\pm0.03$&0.035$\pm0.015$&16.69$\pm11.97$&  5.70$\pm0.01$&0.035$\pm0.015$&$<0.01$\\
6.22&0.030&6.20$\pm0.04$&0.050$\pm0.013$&12.64$\pm7.24$&6.22$\pm0.01$&0.048$\pm0.007$&27.89$\pm6.11$&6.22$\pm0.01$&0.048$\pm0.006$&53.32$\pm10.64$&  6.22$\pm0.01$&0.030$\pm0.008$&19.41$\pm2.35$\\
6.69&0.070&6.69$\pm0.15$&0.070$\pm0.007$&9.31$\pm6.87$&6.69$\pm0.06$&0.070$\pm0.010$& 15.69$\pm7.56$&6.69$\pm0.06$&0.070$\pm0.016$&46.87$\pm19.06$&  6.69$\pm0.01$&0.070$\pm0.007$&7.65$\pm2.97$\\
7.42$^a$&0.126&7.42$\pm0.01$&0.126$\pm0.005$&1.74$\pm3.98$&7.42$\pm0.01$&0.126$\pm0.002$&21.34$\pm5.08$&7.42$\pm0.01$&0.126$\pm0.002$&142.61$\pm11.41$&  7.42$\pm0.01$&0.126$\pm0.008$&7.06$\pm0.80$\\
7.60$^a$&0.044&7.61$\pm0.01$&0.050$\pm0.012$&20.07$\pm0.97$&7.62$\pm0.03$&0.035$\pm0.008$&26.59$\pm4.62$&7.60$\pm0.01$&0.044$\pm0.020$&$<0.01$&  7.57$\pm0.01$&0.040$\pm0.004$&41.75$\pm5.94$\\
7.85$^a$&0.053&7.80$\pm0.01$&0.078$\pm0.011$&26.36$\pm1.35$&7.90$\pm0.02$&0.077$\pm0.005$&55.77$\pm2.27$&7.86$\pm0.01$&0.077$\pm0.003$&146.80$\pm3.01$&  7.86$\pm0.02$&0.053$\pm0.004$&35.21$\pm2.35$\\
8.33&0.052&8.20$\pm0.01$&0.050$\pm0.025$&2.75$\pm2.37$&8.29$\pm0.02$&0.025$\pm0.010$&3.72$\pm3.84$&8.33$\pm0.01$&0.035$\pm0.003$&6.74$\pm1.10$&  8.33$\pm0.01$&0.030$\pm0.006$&16.07$\pm1.46$\\
8.61&0.039&8.59$\pm0.02$&0.040$\pm0.006$&14.16$\pm0.79$&8.60$\pm0.01$&0.075$\pm0.002$&27.46$\pm0.78$&8.61$\pm0.01$&0.080$\pm0.010$&69.81$\pm2.76$&  8.61$\pm0.01$&0.039$\pm0.005$&17.54$\pm1.52$\\
10.68&0.020&10.68$\pm0.08$&0.020$\pm0.010$&$<0.01$&10.68$\pm0.02$&0.020$\pm0.010$&$<0.01$&10.68$\pm0.02$&0.020$\pm0.010$&$<0.01$&10.68$\pm0.05$&0.020$\pm0.010$&$<0.01$\\
11.23$^b$&0.012&11.23$\pm0.01$&0.012$\pm0.002$&9.13$\pm0.05$&11.23$\pm0.01$&0.012$\pm0.002$&10.71$\pm0.36$&11.23$\pm0.01$&0.012$\pm0.002$&20.65$\pm0.28$&  11.23$\pm0.01$&0.012$\pm0.002$&23.12$\pm0.11$\\
11.33$^b$&0.032&11.33$\pm0.01$&0.032$\pm0.002$&37.75$\pm0.22$&11.33$\pm0.01$&0.032$\pm0.003$&46.76$\pm0.41$&11.33$\pm0.01$&0.032$\pm0.001$&112.11$\pm1.06$& 11.33$\pm0.01$&0.032$\pm0.002$&40.85$\pm0.92$\\
11.99&0.045&11.98$\pm0.03$&0.025$\pm0.007$&7.45$\pm1.68$&11.99$\pm0.02$&0.045$\pm0.003$&15.32$\pm2.04$&11.99$\pm0.01$&0.034$\pm0.004$&43.00$\pm4.19$&  11.97$\pm0.01$&0.025$\pm0.005$&13.82$\pm2.00$\\
12.62$^c$&0.042&12.62$\pm0.1$&0.047$\pm0.004$&19.53$\pm0.39$&12.62$\pm0.01$&0.047$\pm0.003$&23.42$\pm1.64$&12.62$\pm0.01$&0.047$\pm0.002$&63.49$\pm4.11$&  12.62$\pm0.01$&0.047$\pm0.006$&27.32$\pm1.32$\\
12.69$^c$&0.013&12.69$\pm0.01$&0.013$\pm0.004$&1.87$\pm1.12$&12.69$\pm0.01$&0.013$\pm0.002$&2.84$\pm1.20$&12.69$\pm0.01$&0.013$\pm0.002$&6.40$\pm2.01$&  12.69$\pm0.01$&0.013$\pm0.002$&4.06$\pm1.62$\\
13.48&0.040&13.48$\pm0.04$&0.070$\pm0.014$&7.20$\pm0.85$&13.43$\pm0.15$&0.070$\pm0.009$&15.72$\pm1.56$&13.48$\pm0.04$&0.070$\pm0.012$&59.64$\pm2.20$&  13.50$\pm0.08$&0.070$\pm0.011$&17.08$\pm1.08$\\
14.04&0.016&14.04$\pm0.05$&0.016$\pm0.008$&$<0.10$&14.04$\pm0.17$&0.016$\pm0.010$&$<0.10$&14.04$\pm0.09$&0.016$\pm0.008$&$<0.01$&  14.04$\pm0.10$&0.016$\pm0.009$&$<0.01$\\
14.19&0.025&14.40$\pm0.04$&0.025$\pm0.008$&1.45$\pm0.69$&14.40$\pm0.06$&0.050$\pm0.015$&10.05$\pm3.11$&14.40$\pm0.04$&0.050$\pm0.008$&31.71$\pm3.47$&  14.40$\pm0.07$&0.050$\pm0.009$&3.31$\pm0.91$\\
15.90&0.020&15.90$\pm0.04$&0.015$\pm0.005$&1.67$\pm0.08$&16.05$\pm0.01$&0.030$\pm0.004$&3.65$\pm0.18$&16.10$\pm0.07$&0.030$\pm0.012$&10.96$\pm0.49$&  15.90$\pm0.01$&0.020$\pm0.004$&5.51$\pm0.30$\\
16.45$^d$&0.014&16.39$\pm0.01$&0.014$\pm0.003$&3.51$\pm0.08$&16.55$\pm0.01$&0.014$\pm0.001$&2.35$\pm0.65$&16.45$\pm0.01$&0.015$\pm0.003$&8.76$\pm0.45$&  16.40$\pm0.01$&0.015$\pm0.003$&8.16$\pm0.35$\\
17.04$^d$&0.065&17.04$\pm0.01$&0.065$\pm0.002$&16.95$\pm0.15$&17.04$\pm0.05$&0.065$\pm0.001$&18.46$\pm0.20$&17.04$\pm0.01$&0.065$\pm0.004$&56.02$\pm0.87$&  17.04$\pm0.01$&0.065$\pm0.003$&2.57$\pm0.40$\\
17.37$^d$&0.012&17.37$\pm0.04$&0.020$\pm0.005$&2.79$\pm0.56$&17.37$\pm0.02$&0.012$\pm0.002$&1.84$\pm 0.51$&17.37$\pm0.01$&0.012$\pm0.009$&2.78$\pm0.70$&17.30$\pm0.02$&0.025$\pm0.002$&9.54$\pm0.72$\\
17.87$^d$&0.016&17.87$\pm0.18$&0.016$\pm0.003$&0.18$\pm0.07$&17.70$\pm0.02$&0.016$\pm0.003$&2.34$\pm0.18$&18.10$\pm0.02$&0.016$\pm0.003$&1.70$\pm0.17$&  17.86$\pm0.01$&0.025$\pm0.003$&1.93$\pm0.41$\\
18.92&0.019&18.70$\pm0.25$&0.010$\pm0.012$&0.54$\pm0.04$&18.70$\pm0.04$&0.039$\pm0.005$&2.10$\pm0.80$&18.70$\pm0.01$&0.039$\pm0.006$&13.20$\pm0.64$&  18.70$\pm0.01$&0.039$\pm0.004$&3.31$\pm0.56$\\
&&19.15$\pm0.04$&0.030$\pm0.010$&2.05$\pm0.90$&19.10$\pm0.01$&0.030$\pm0.010$&0.99$\pm 0.22$&19.10$\pm0.03$&0.030$\pm0.007$&6.52$\pm1.20$&19.10$\pm0.01$&0.030$\pm0.015$&$<0.01$\\

\hline
\end{tabular}

Notes: $^{*}$Central wavelength ($\lambda_{\rm{0}}$) and fractional FWHM
($\gamma$) of the features from DL07. $^\natural$ $\lambda_{\rm{0}}$,
$\gamma$,  and integrated fluxes, $F$, of each feature given by our
best fit to the MIR spectra. $^{(a)}$ 7.7 $\mu$m complex. $^{(b)}$
11.3 $\mu$m complex. $^{(c)}$ 12.7 $\mu$m complex. $^{(d)}$ 17
$\mu$m complex
 $^{(e)}$ Units in $\mu$m; $^{(f)}$ Units in
$10^{-18}$ W m$^{-2}$ \label{table3}
\end{table*}
\end{turnpage}

The flux calibrated de-redshifted IRS spectra of NGC~1297,
NGC~5044, NGC~6868, and NCG~7079 are shown in Figure \ref{fig1}.
In the same figure we also plot, for comparison, the spectra of an ETG which has experienced a recent
episode of star formation,  NGC~4435 \citep{panuzzo07}, and of two typical passively evolving
ETGs, the cluster ETG NGC~4621 \citep[see][]{bressan06} and the field ETG NGC~3818 \citep{Panuzzo10}.
In spite
of the similarity to the  passive galaxies, especially in the lower
wavelength spectral range, the IRS spectra of our objects show
several signatures of activity, such as: (a) strong PAH emission at
$\lambda_{\rm{0}} \sim 11.3$ $\mu$m,
 similar to that found by \citet{kaneda05} in four ETGs and the well developed
complex of PAH emission near  $17$ $\mu$m;  (b) other weak dust
features at 6.2 $\mu$m, 7.7 $\mu$m, 8.6 $\mu$m, and 12.7 $\mu$m, usually attributed to PAH molecules \citep[e.g.][]{allamandola89,puget89,tielens08} ; (c)
 strikingly strong emission of pure H$_2$ rotational lines; (d)
 forbidden nebular emission lines of Ar, Fe, Ne, O, S, and Si;
and (e) warm dust continuum emission.

\section{Analysis of the PAH emission features}
\label{sec:method}

\subsection{Removal of the underlying stellar and dust continua}

In order to  study the PAH features it is necessary to properly remove  from each spectrum
the underlying stellar and the dust continua.

We adopt a semi-empirical, high S/N, template derived from the analysis of  passive
ETGs to describe the stellar continuum representative of the old stellar population.
That template was built averaging the NIR (J-H-K 2MASS) data and
the 5--40 $\mu$m IRS-Spitzer spectra of three passively evolving ETGs in low density
environments, namely NGC~1389, NGC~1426, and  NGC~3818 \citep{Panuzzo10}.
The  stellar continuum  template shows the dip at 8~$\mu$m likely due to
photospheric SiO absorption bands \citep[e.g.][]{ver09} and
the bump of silicate emission at $\approx$10~$\mu$m likely due to dusty circumstellar
O-rich AGB envelopes. Both  are features typical of passively evolving ETGs
\citep{bres98,bressan06}. In Figure~\ref{fig1}, as an example, we illustrate the range of variation
of these stellar features in the stellar continuum of  NGC~4621 and NGC~3818
located in Virgo and in the field, respectively.

In order  to properly remove the underlying continuum from the IRS spectra, we use the 2MASS
magnitudes within an aperture of 5\arcsec\ radius, which is close to that of our IRS spectra
\citep{panuzzo07}. The NIR magnitudes and errors are listed in Table~\ref{tablenir}.  Assuming
that the NIR fluxes
  are completely due to the stellar component,  we normalize our stellar
  continuum template to the observed flux in the H-band, and calculate the contribution of the stellar
  continuum to the MIR spectra. We find the old stellar population contributes about $97.5\%$, $98.5\%$, $99.6\%$, and $98.5\%$ of the observed fluxes at 5.6 $\mu$m for NGC~1297, NGC~5044, NGC~6868,
  and NGC~7079, respectively. We then may consider that the old stellar population dominates the
  MIR emission at shorter wavelengths. This was already
  expected from  the comparison with passive ETGs
  shown in Fig.~\ref{fig1}. However, this is not always the case. Indeed Panuzzo et al. (2007) found that
  the contribution of the stellar component to the MIR emission of NGC 4435 was less than 30$\%$ at $5.6\;\rm \mu m$.


After a proper subtraction of the stellar  continuum, the galaxies show a residual continuum
component due to dust emission. It is removed by fitting up to 5 modified
black-bodies, with temperatures ranging from 20 to 350 K.
Notice that these dust components are chosen just to smoothly
reproduce a realistic underlying dust continuum without any
attempt to provide a physical interpretation.

The whole process of continuum subtraction and the final results are shown
 in Figure~\ref{fig2}. The final spectra are  a combination of emission lines and dust features.

 All the spectra, normalized to the peak of the 11.3~$\mu$m feature, are shown
in Figure \ref{fig5}. For comparison, we also plot in the same figure the spectrum
of  NGC~1482, a HII galaxy
with PAH emission typical of a star-forming galaxy  \citep{smith07}.

\subsection{Analysis of the dust feature emission}
\label{sec:results}

Before discussing the individual PAH features, we point out the following  points
on the general shape of the spectra:

\begin{itemize}

\item The continuum subtracted spectra of our ETG sample are very similar. At the same time,
 in the 5--9 $\mu$m spectral range, they significantly differ from star forming galaxies, represented by
 NGC~1482.

\item
A broad 7.7~$\mu$m feature, hardly recognizable
in the original spectra,  appears after continuum subtraction because of
the presence of the dip at 8~$\mu$m in the underlying
stellar continuum.

\item The PAH bands at 8.6, 7.7, and primarily at 6.2~$\mu$m are much less intense
than those  observed in the integrated spectrum of star forming galaxies (see the NGC~1482 spectrum).
On the contrary, the emission at 17~$\mu$m looks very similar to that of
star forming galaxies.

\item The positions of the peak of the 7.7~$\mu$m band in NGC~6868 and in
NGC~5044, are shifted to longer wavelengths.

\item Despite the similarity of  the PAH emission bands in the 4 ETGs,
the emission lines show considerable variation from galaxy to galaxy.
These emission lines will be analyzed for a larger sample in a companion paper
\citep{Panuzzo10}.
\end{itemize}

\begin{table*}
\centering
\caption{PAH interband ratios and  PAH to continuum ratios}
\begin{tabular}{l|cccc|c}
\hline\hline
&{N1297}&{N5044}&{N6868}&{N7079}&SINGs median value$^{(b)}$\\
\hline
6.2/7.7$^a$&$0.20\pm0.11$&$0.15\pm0.06$&$0.18\pm0.04$&$0.23\pm0.03$&0.28 [0.21 -- 0.34]\\
6.2/11.3$^a$&$0.27\pm0.10$&$0.47\pm 0.11$&$0.40\pm0.09$&$0.30\pm0.04$&1.10 [0.77 -- 1.43]\\
6.2/17$^a$&$0.54\pm0.31$&$0.51\pm0.21$&$0.77\pm0.15$&$0.87\pm0.11$&1.90 [1.33 -- 2.47]\\
7.7$^a$/11.3$^a$&$1.36\pm 0.09$&$1.80\pm0.13$&$2.18\pm 0.09$&$1.26\pm 0.10$&3.60 [2.52 -- 4.68]\\
7.7$^a$/8.6&$3.40\pm 0.36$&$3.77\pm 0.28$&$4.14\pm0.23$&$4.79\pm 0.55$&5.70 [3.99 -- 7.41]\\
7.7$^a$/17$^a$&$2.72\pm 0.20$&$3.34\pm 0.26$&$4.18\pm0.18$&$3.78\pm 0.33$&6.90 [4.83 -- 8.97]\\
8.6/11.3$^a$&$0.31\pm0.04$&$0.47\pm0.03$&$0.53\pm0.02$&$0.27\pm0.02$&0.68 [0.48 -- 0.88]\\
11.3$^a$/12.7$^a$&$2.19\pm 0.12$&$2.18\pm 0.17$&$1.90\pm 0.13$&$2.03\pm 0.14$&1.80 [1.26 -- 2.39]\\
17$^a$/11.3$^a$&$0.50\pm 0.01$&$0.44\pm 0.02$&$0.52\pm 0.01$&$0.35\pm 0.02$&0.53 [0.37 -- 0.69]\\

\hline
L$_{\rm{{PAH}}}(6.2)/\nu L_\nu(24)$&0.058$\pm 0.034$&0.039$\pm 0.009$&0.044$\pm 0.010$&0.031$\pm 0.005$&0.10 [0.05 -- 0.20]\\
L$_{\rm{{PAH}}}(7.7)/\nu L_\nu(24)$&0.220$\pm 0.029$&0.148$\pm 0.010$&0.240$\pm 0.025$&0.132$\pm 0.017$&0.40  [0.25 -- 1.00]\\
L$_{\rm{{PAH}}}(11.3)/\nu L_\nu(24)$&0.215$\pm 0.022$&0.082$\pm 0.008$&0.110$\pm 0.011$&0.101$\pm 0.010$&0.10 [0.04 -- 0.20]\\
L$_{\rm{{PAH}}}(17)/\nu L_\nu(24)$&0.107$\pm 0.011$&0.038$\pm 0.004$&0.057$\pm 0.006$&0.035$\pm 0.004$&0.06 [0.03 -- 0.10]\\
\hline

 \hline
\end{tabular}

Notes:  $^a$
blended PAH complex, see Table \ref{table3};
$^{(b)}$ median value  and 1 $\sigma$  variation in the
corresponding ratio derived by  \citet{smith07} for the SINGs galaxy sample. \label{table4}
\end{table*}

\subsubsection{The individual features}

We use a spectral decomposition method similar to that outlined by \citet{smith07}
to recover the PAH features and the emission lines. In this method, the spectrum is modelled by a combination of Gaussian profiles for  the atomic and molecular emission lines, and Drude profiles for the PAH features \cite[e.g.][]{ld01}.
The broad PAH features are  represented by up to four sub-features grouped as blended complexes, some of them adopted to best reproduce the band shape (e.g. Smith et al. 2007). In order to explore if the peculiarities of our PAH spectra could be due not only to differences in the relative intensities of the sub-features, but also to differences in their characterizations, we leave as free parameters the central wavelengths,
$\lambda_{\rm{0}}$ and the fractional full width at half maxima (FWHM), $\gamma$,
of the Drude profiles. The model is then fitted to the observed spectra by minimizing the global $\chi^2$  using the Levenberg-Marquardt algorithm, and using the values $\lambda_{\rm{0}}$  and $\gamma$ reported by \citet{Draine07} (hereafter DL07) as the initial guess for the parameters of the Drude profile.

The best fits to the spectra are shown as thick red lines  in Figure
\ref{fig3}, where the fitted PAH subfeatures and emission lines are shown as thin green and thin orange lines, respectively.
The results of the fits are summarized in Table \ref{table3}.
Columns 1 and 2 list the central wavelengths,
$\lambda_{\rm{0}}$ and the fractional FWHM, $\gamma$, reported by  DL07.

In the remaining columns we report, for each galaxy, the fitted
$\lambda_{\rm{0}}$,  $\gamma$, and the corresponding integrated fluxes.
Different sub-features of the 7.7 $\mu$m,
11.3 $\mu$m, 12.7 $\mu$m, and 17 $\mu$m PAH complexes,
are indicated with the letters $a$, $b$, $c$, and $d$, respectively.
The errors in the derived fluxes and Drude profile parameters have been estimated
repeating the above procedure with simulated spectra
obtained by randomly varying the fluxes within the observational error.
The process was repeated 100 times and the errors were calculated as
the standard deviation of the distribution of the fitted values in
the simulations.

From Table \ref{table3} we notice that even though most of the $\lambda_0$ and  $\gamma$
of the fitted features are in good agreement with those reported by DL07,
the fits for NGC~6868, NGC~5044,
and to a  lesser extent NGC~1297,
required some
Drude profiles with significantly different lambda peaks and/or fractional FWHM.
For some  features the differences may be due to
low signal to noise (i.e. 5.27 $\mu$m) but this is not the case for the 7--9 $\mu$m
complex, which is well detected in all cases.

The most remarkable differences with respect to DL07 are:
\begin{itemize}
\item The weakness or even lack  of the 7.6 $\mu$m feature in
NGC~5044 and NGC~6868. The 7.6 $\mu$m feature is the
main component of the 7.7~$\mu$m  complex in normal and active
galaxies \citep[e.g.][]{peeters04,smith07,tielens08}.
\item The brightest component is the 7.85~$\mu$m
sub-feature.  It is also about twice as wide ($\gamma \sim
0.077$) as that in DL07.
Moreover, in the case of NGC~5044, it peaks at $\lambda_{\rm{0}} = 7.90 \mu$m.

\item The fits to the broad  6--9 $\mu$m feature in the NGC~5044 and NGC~6868 spectra require also of 2 times
 wider ($\gamma \sim 0.75-0.8$)  Drude profiles for the 8.6 $\mu$m feature.

\item The 17~$\mu$m PAH complex is also wider than in normal galaxies and
requires an ``extra" feature at about 19.1 $\mu$m.
\end{itemize}

The fitting requirements of wider Drude profiles
for the well established features at 7.7 $\mu$m and 8.6 $\mu$m are actually due to an emission excess in the 6--9 $\mu$m spectral range with respect to what is recovered by  using only the \textit{standard} PAH Drude profiles (i.e. DL07's features). Thus, the broad 6--9 $\mu$m feature in NGC~5044 and NGC~6868 could also be well reproduced
by adding some \textit{extra} Drude profile sub-features to the \textit{standard} PAH feature set. 

In order to test if  this \textit{excess} emission could be due to the use of a different decomposition method to that used by Smith et al. (2007), we applied our procedure to NGC~4435
\citep{panuzzo07}  and to NGC~1482 \citep{smith07} from the Smith et al. sample.
In neither case did we find significant differences with respect to
the values reported by DL07.
By using the  NGC~1482 spectrum we also test the dependence of our results on the different methods
of subtraction of the underlying continuum. Following the recipe described in Section~3
we find that the contribution from the old stellar population to the 5.6 $\mu$m flux is $< 28 \%$, and that the differences between the continuum subtracted spectrum obtained with our method and with PAHFIT (Smith et al. 2007) are negligible. Thus, we are confident that our results are not an artifact of the
adopted method for the subtraction of the underlying continuum.
Notice however, that for a correct subtraction of the stellar component, the use of a proper stellar template
is mandatory for  ETGs at odds with HII galaxies where the contribution of the old stellar population
 to the MIR spectral range is small.

\subsubsection{PAH inter-band ratios}

The strengths of the individual PAH bands are affected by a variety of mechanisms, that include the distribution of PAH sizes and charges, PAH de-hydrogenization, and the strength and hardness of the radiation field exciting the PAHs \citep[e.g.][]{tielens08,galliano08}. Possible destruction mechanisms of PAHs are also varied, including photo-dissociation, sputtering and dissociation in shocks \citep[e.g.][]{leger89,micelota1,micelota2}.  Variations of the PAH inter-band ratios can be explained by the combined effects of those  mechanisms, and hence to variations in the  physical conditions of the environment \citep[e.g.][]{allamandola89,schutte93,Draine07,galliano08,tielens08}. 

Table \ref{table4} lists the ratios of the strongest PAH bands.
In the last column of the table we report the median values and their  1 $\sigma$
variation for the 59 objects of the SINGs galaxy sample \citep{smith07}.
The most striking characteristic of the spectra is the
unusually low 7.7/11.3 $\mu$m PAH inter-band ratio
with respect to what is observed in star forming galaxies.
This  unusual ratio  was already reported by
\citet{kaneda05} and \citet{bressan06} in a small set of ETGs and later confirmed by \citet{kaneda08} for a larger sample of ETGs  and by \citet{smith07} for a subset of about 10 objects in the SINGs galaxy sample. A more careful inspection of the spectra indicates that the 7.7/11.3 $\mu$m PAH inter-band ratio is not alone in presenting these unusual values but
there is a systematic deficit  in the shorter wavelength bands (i.e. 6.2, 7.7, and 8.6 $\mu$m) with respect to those peaking at longer wavelengths (i.e. 11.3, 12.7, and 17 $\mu$m).
This effect is more extreme at shorter wavelengths, with the 6.2 $\mu$m band having the strongest deficit. The average 6.2/11.3 ratio for our four galaxies is $0.36\pm0.20$
which is significantly different from $1.10\pm0.33$ of the
 SINGs galaxies. The same for 6.2/17, our is $0.67\pm0.30$ while
SINGs is $1.9\pm0.57$. Differences in the
7.7/11.3 and 7.7/17 $\mu$m ratios are less extreme. Our average 7.7/11.3 and 7.7/17 $\mu$m ratios are $1.65\pm0.54$ and $3.51\pm0.80$, while the mean values for the SINGs sample are $3.60\pm1.08$ and $6.90\pm2.07$, respectively.
 In the case of the 8.6 $\mu$m band, only  NGC~1297 and NGC~7079 present 8.6/11.3 $\mu$m and 8.6/11.3 $\mu$m ratios which are significatively lower than the values of the SINGs sample. Whereas,  17/11.3 $\mu$m and the 12.7/11.3 $\mu$m ratios are, in all the cases, within the  values found for the SINGs galaxy sample. 
We check wether these differences could be due to some general PAH emission defect
with respect to dust emission. In the last rows of Table \ref{table4} we report
the ratio between PAH complexes and the MIPS$_{\rm{24\mu m}}$ luminosity.
The last column also provides the median values of the SINGs sample.
We see again that the deficit is selective, and it occurs only in the shorter wavelength bands,
while at 11.3$\mu$m and 17$\mu$m, the ratios fall within those of the SINGs sample.

Laboratory studies and theoretical calculations show that
the $6-9\mu$m bands are intrinsically weak in neutral PAHs and become
stronger when the PAHs are ionized (Allamandola
et al. 1999; Bauschlicher 2002; Kim \& Saykally 2002). Therefore,
the  $6-9\mu$m bands will be much more intense for an ionized PAH
than for a neutral PAH, while the contrary will be true for the 3.3 and
11.3 $\mu$m bands. Consequently, the ratios between the $6-9\mu$m bands and the
11.3 $\mu$m band depend on the charge of the PAHs, which
is directly related to the physical conditions (intensity of the ionizing
radiation field, electron density, etc.) of the PAH emitting region. On the other hand, larger PAHs attain lower temperatures than their smaller counterparts upon the absorption of the same photon. That implies that  smaller PAHs will contribute mainly at short wavelengths, while larger PAHs account mainly for  the emission at long wavelengths \citep[e.g.][]{schutte90,schutte93,Draine07,bau08,bau09,boe10}. \citet{boe10} found that while the $6-9$ $\mu$m and 11.3 $\mu$m bands are very sensitive to the ionization state of the PAH molecule, the 17 $\mu$m complex is not, and that the emission of this complex is probably due to the C--C--C vibrational modes of a limited number of large PAHs.
Consistently, the ratios of the 17/$6-9$ $\mu$m features would be a sensitive probe of the size of interstellar PAHs so that size distributions biased to large PAHs would result in higher values of the 17/6.2 $\mu$m and 17/7.7 $\mu$m ratios compared to  the values in normal galaxies.
On the other hand, charge distributions biased to neutral PAHs will result in lower 7.7/11.3 $\mu$m and 6.2/11.3 $\mu$m ratios compared to those in normal galaxies.

Therefore, the low  values of the 6.2/17 $\mu$m  and the 7.7/17 $\mu$m ratios in our ETGs imply a size distribution deficient in small PAHs, and responsible for the "normal" PAH emission at 17 $\mu$m. This suggestion is reinforced by the small values of the 6.2/7.7 $\mu$m ratio which is particularly sensitive to the lower cutoff of the PAH size distribution \citep[e.g.][]{Draine07,galliano08}. However, theoretical computations of large ionized PAH spectra indicate an enhanced emission of the $6-9\mu$m bands with respect to the 11.3 $\mu$m band, even if the effects temperature cascade is taken into account in their computations \citep[see Figure 17 of][]{boe10}. Therefore, a mixture of large PAHs with a \textit{normal} charge distribution could not reproduce the low values of the 7.7/11.3 $\mu$m ratios observed in our objects, which would require of a mixture plentiful of neutral PAHs. In summary, the low values of the $6-9 \mu$m bands versus the 11.3 $\mu$m and the 17 $\mu$m bands observed in our objects are compatible with the emission from a mixture which is rich in large and neutral PAHs.

\smallskip

\section{Characterization of the PAH emission}


Instead of using  combinations of single Drude profiles,  we will now
analyze the continuum subtracted spectra by means of {\it emission templates}
of different dust components. We will follow the method
devised by \citet{rapacioli05}, \citet{berne07}, \citet{joblin08},
\citet{joblin09} and \citet{berne09} which provides general information on the
sizes and ionization states of PAHs, and the presence of other dust components.
In this way, we will gain insight into the nature of these spectral features
and, ultimately, about the stellar populations and physical processes involved.
Before using the emission templates, we removed the emission lines from the
continuum subtracted spectra by using the values of the lines derived
in the fits performed in section \ref{sec:results}.

\subsection{The Templates}

\begin{figure}
\includegraphics[width=0.5\textwidth]{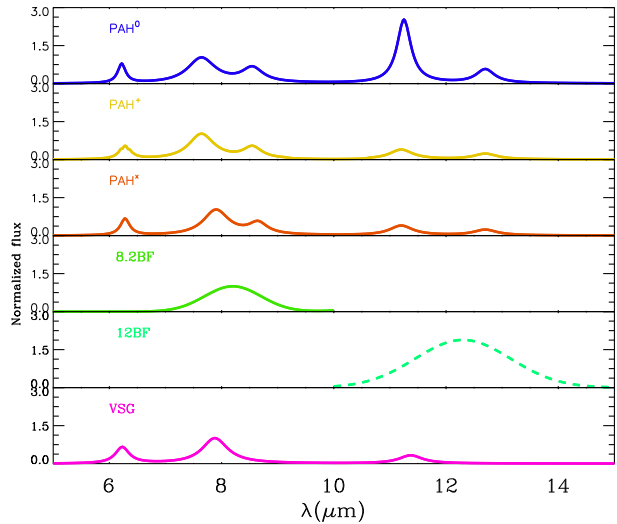}
\caption{Templates from \citet{joblin08}. Fluxes are normalized
to the peak of the 7.7 $\mu$m component.}
\label{fig7}
\end{figure}

There are several studies in the literature about emission templates of PAH families
and other dust features
\citep[e.g.][]{cesar00,rapacioli05,rapacioli06,berne07,joblin08,joblin09,berne09}.
For the following analysis we will use the PAH--dust feature
templates from \citet{joblin08} which are shown in Figure \ref{fig7}.
The parameters of the templates (i.e. $\lambda_{\rm{0}}$ and FWHM of the Lorentzian and Gaussian profiles)
were taken from their Table A.1.
Three of them were obtained from the mathematical decomposition of three
reflection nebulae  observed with ISO and \emph{Spitzer} by using the Principal Component Analysis technique \citep{rapacioli05,berne07}.
The first template, called PAH$^+$, characterized by strong
C--C modes (6 -- 9 $\mu$m) relative to C--H
modes (10 -- 14 $\mu$m) and peaking at 7.6~$\mu$m, is attributed to PAH cations \citep{rapacioli05}.
The second template called PAH$^0$, characterized by
strong C--H modes relative to C--C modes and showing an evident peak at
11.3~$\mu$m, is attributed by the authors to a PAH mixture rich in neutral PAHs \citep{rapacioli05}.
The third one called VSG consists of weaker and broader features and is likely due to
PAH clusters (Cesarsky et al. 2000, Rapacioli et al. 2005, Rapacioli et al. 2006).
The same authors suggest that PAH clusters could break into free PAHs
 under the effects of a UV radiation field.


\begin{figure}
\includegraphics[width=0.49\textwidth]{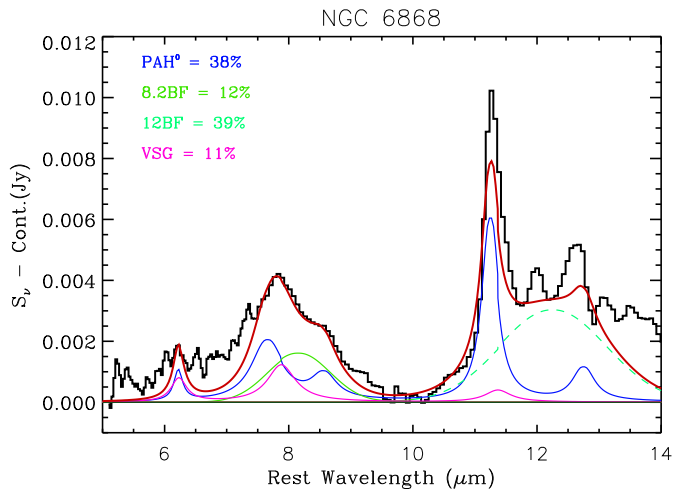}
\caption{Analysis of the PAH spectrum of NGC~6868 after emission line subtraction (solid black line) with the template spectra
displayed in Figure \ref{fig7}. The best fit is shown as a solid red line, while the contribution of
each  PAH component is shown with the same colour code as in Figure
\ref{fig7}. In the legend, we display the percentage contribution of the different PAH components to the
integrated flux in the range between 5.5 and 14 $\mu$m.}
\label{fig8}
\end{figure}
\begin{figure}
\includegraphics[width=0.5\textwidth]{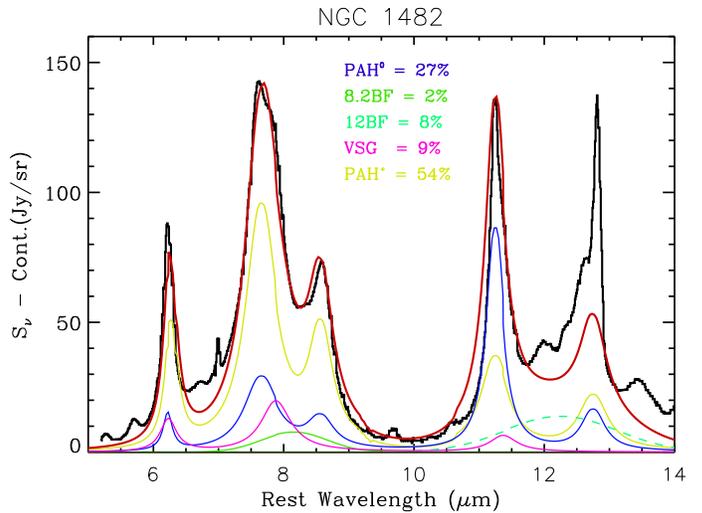}

\caption{Fit to the continuum subtracted spectrum of the  HII galaxy NGC~1482 \citep{smith07} by using the set of nine dust feature
template spectra. The meaning of the lines is as in Figure \ref{fig9}. Notice  that  the main contributor
to the integrated flux is the PAH cations, as is expected in an
object with a  burst of star formation \citep{allamandola99}. The BF contributions are very {\bf small}.} \label{fig10}
\end{figure}
\begin{figure*}
\includegraphics[width=0.5\textwidth]{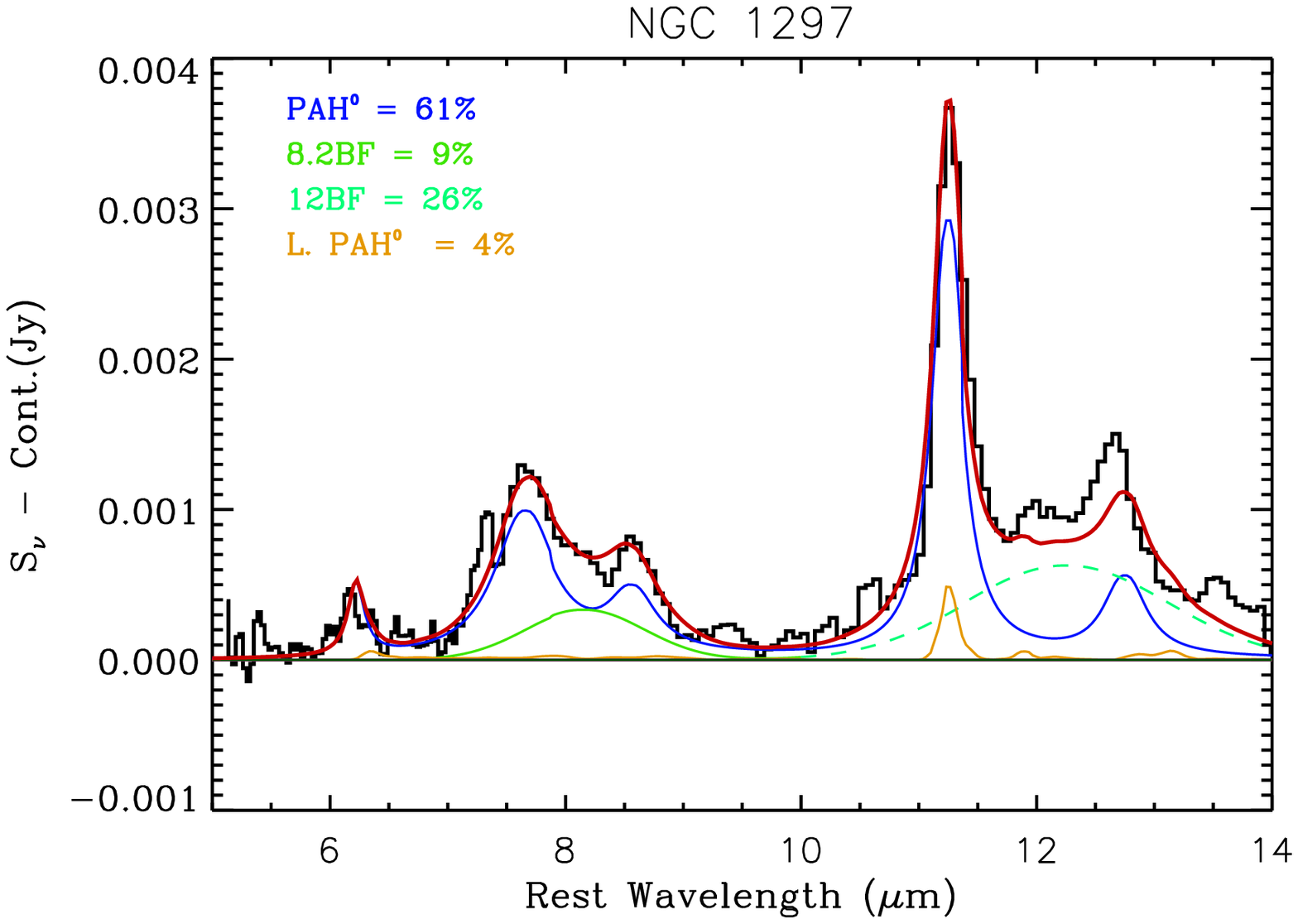}
\includegraphics[width=0.5\textwidth]{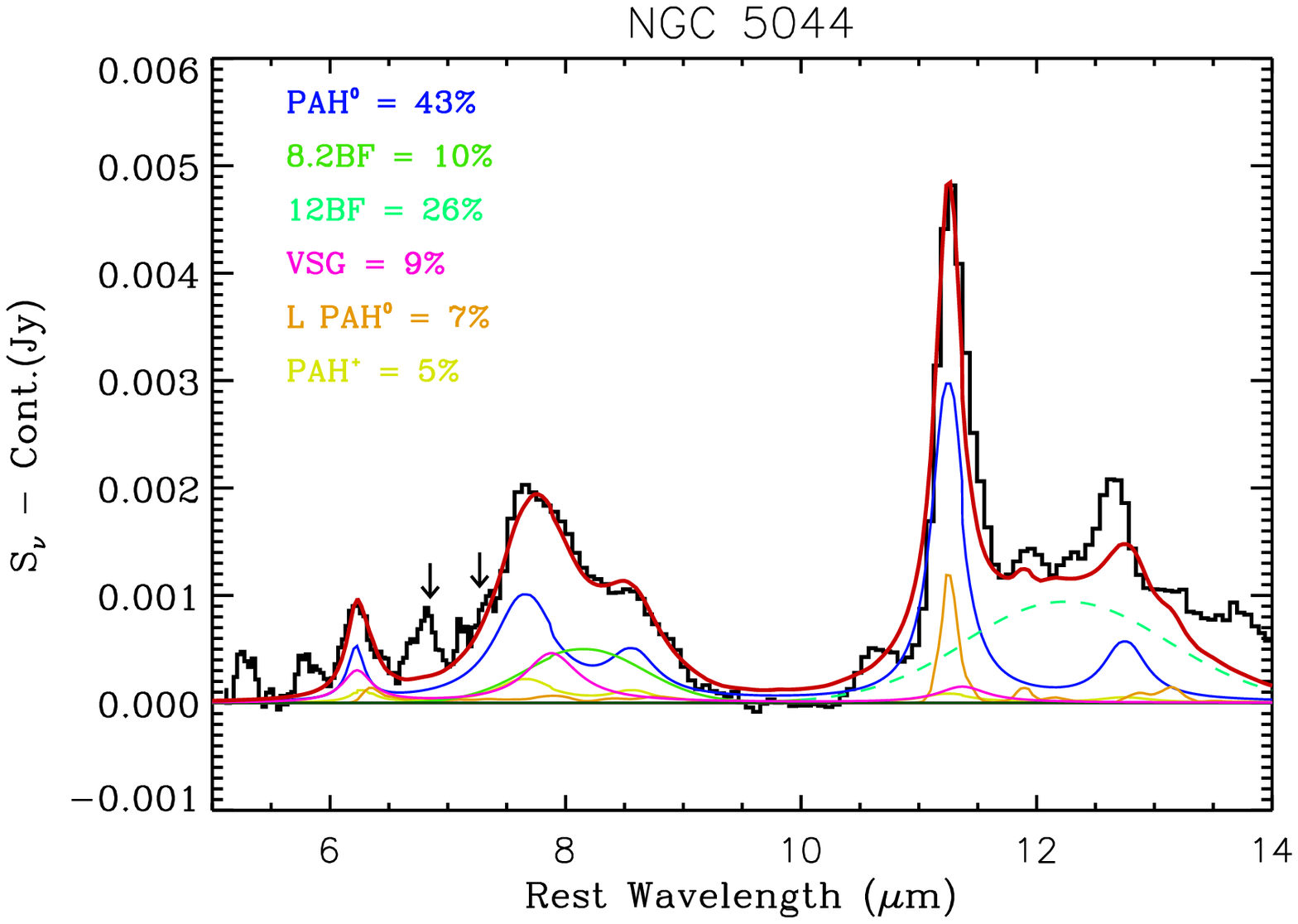}
\includegraphics[width=0.5\textwidth]{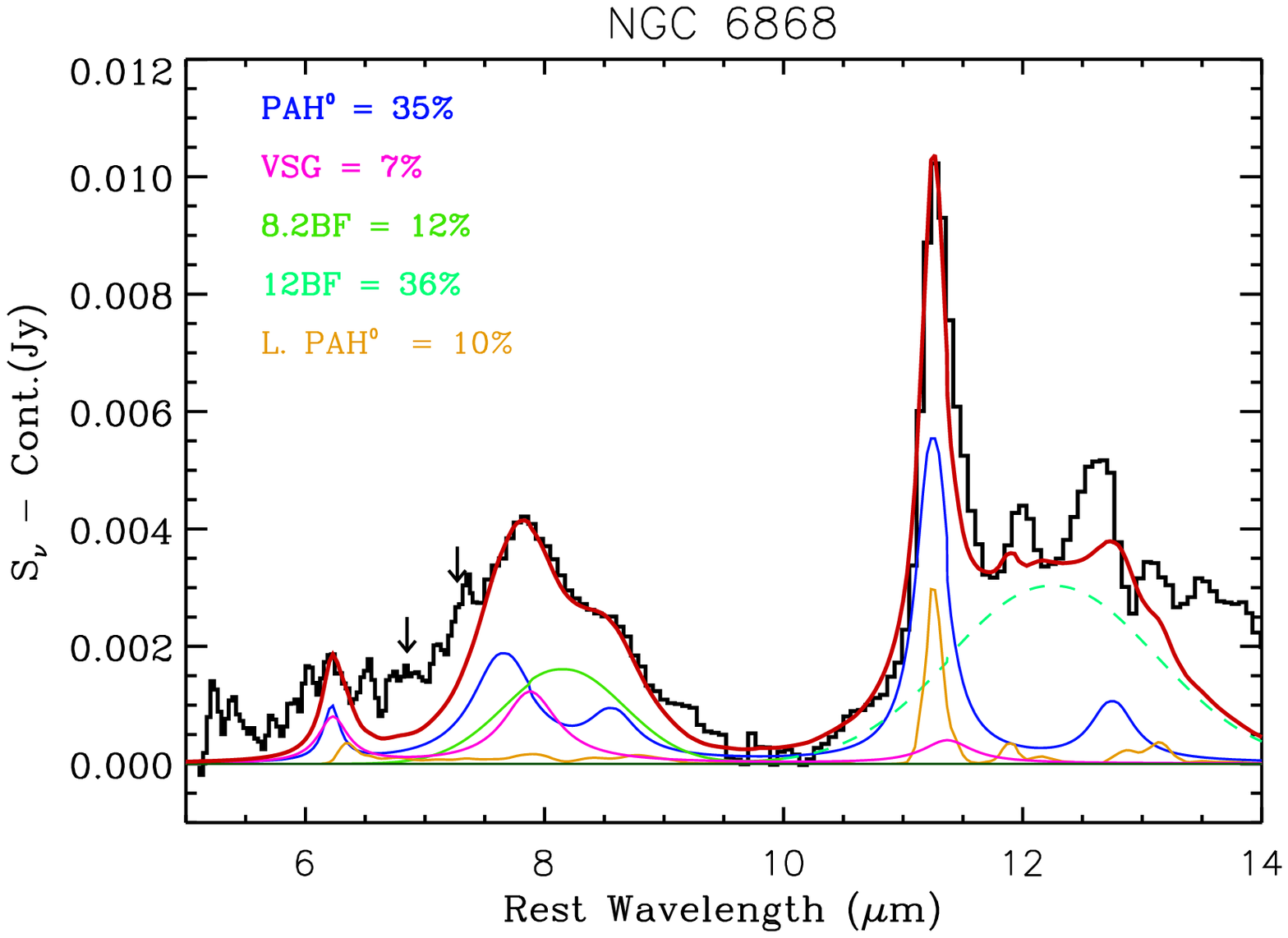}
\includegraphics[width=0.5\textwidth]{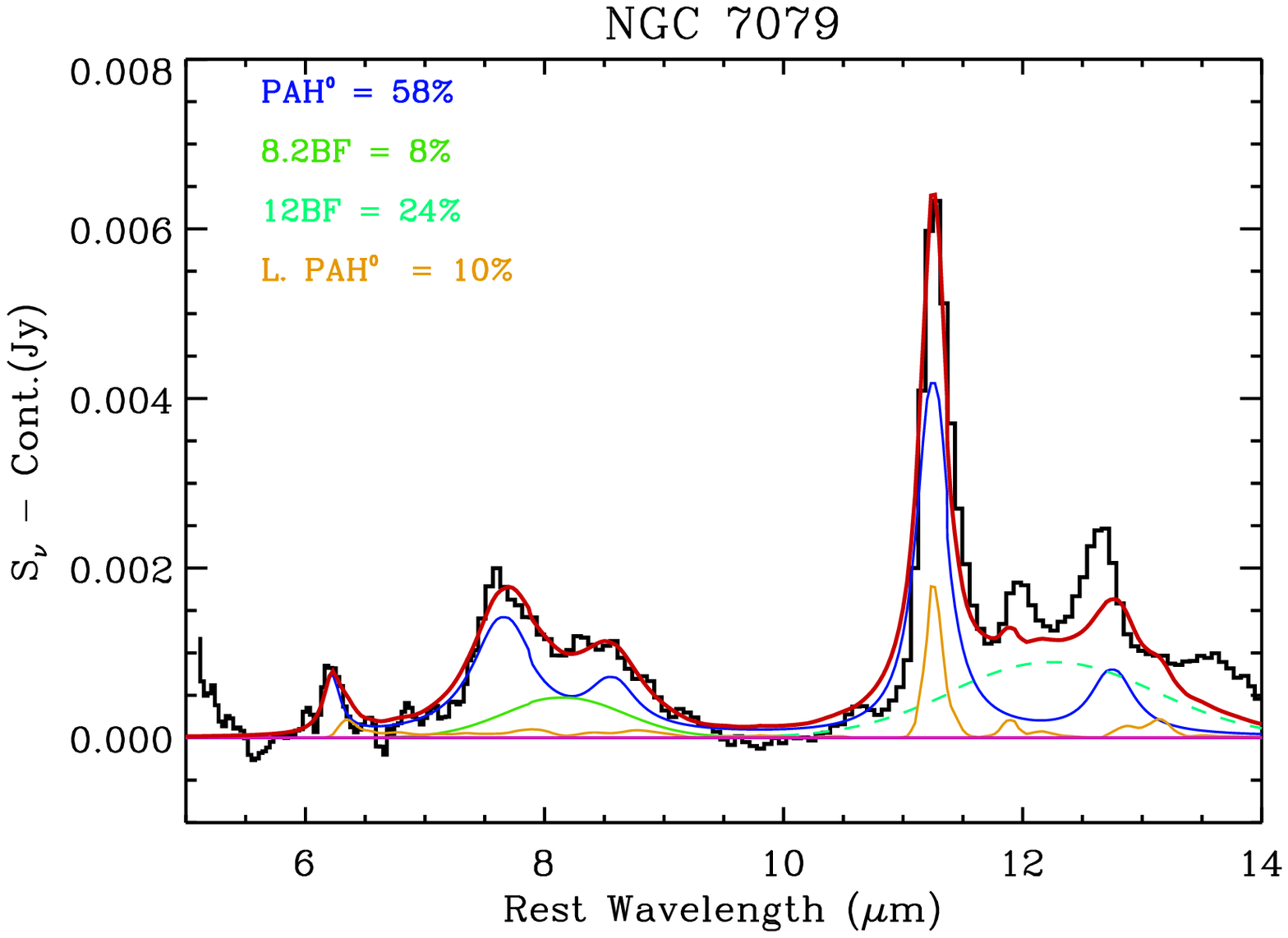}
\caption{Best fits (solid red line) to the PAH spectra of our ETGs
(solid black line). The different  templates are shown in colour:
PAH$^0$(solid blue line), 8.2BF component (solid green line),
12BF component (dashed cyan line), VSG (solid magenta line),
and  Large neutral  PAHs (solid orange line).  In the legend of each panel,
we report the percentage of the relative contribution of the different
dust feature components to the integrated flux in the range 5.5 to
14 $\mu$m. Arrows in the spectra of NGC~5044 and NGC~6868
indicate the possible detections of the 6.85$\mu$m and 7.25$\mu$m
features usually ascribed to aliphatic material \citep[e.g.][]{chiar00,sloan07}.}
\label{fig9}
\end{figure*}

The other three observational components were later introduced by
\citet{joblin08}. They were required by those authors to
fit the spectra of Galactic and Magallanic  Cloud
planetary nebulae (PNe) observed with \emph{Spitzer} and ISO, which could
not be fitted only with the templates derived by \citet{rapacioli05}.
Two of them are broad emission features (BF) at 8.2$\mu$m and 12$\mu$m
already seen in some post-AGB stars
\citep{kwok89,hrivnak00,hony01,peeters02,van04,sz05,kraemer06}.
Peteers et al. (2002) classified as C-class objects those in
which the broad 8.2 $\mu$m feature is observed. The 8.2BF
and the 12BF templates were built by Joblin et al. by fitting those
observed bands in the post-AGB star IRAS 13416-6243 MIR spectrum, which is
classified as a C-class object prototype by Peeters et al. (2002).
It is worth recalling that the contributions of the broad features (i.e. 8.2BF and 12BF)
are important only for the objects that are thought to be in \textit{transition}
from the AGB to the PN phase \citep[e.g.][]{peeters02,van04,sloan07,tielens08,joblin09,berne09}.
The second one is a ``new'' PAH type template spectrum, named {\it PAH$^{\rm{x}}$},
with the 7.8 $\mu$m band shifted
to $7.9 \mu$m and the 8.55 $\mu$m shifted to 8.65 $\mu$m.
This template was introduced ``ad hoc" by \citet{joblin08}, who
tentatively associated it with large (N$_{\rm{C}} \ge 100$) PAHs, likely anions, although
the large cations option could not be excluded \citep{bau08,bau09}.

With the empirical templates, we obtained best fits like that shown
in Figure \ref{fig8} for NGC~6868.
Though the best fit represents fairly well the general shape of the PAH spectrum
we find that it underestimates significantly the flux of the 11.3 $\mu$m feature.
In Figure \ref{fig10} we show the best fit for NGC~1482. In this case, the  MIR spectrum  is  well reproduced with the set of templates, indicating that, at odd of our ETGs, the \textit{usual} PAH spectrum  can be depicted by a variable combination
of these dust components, mainly by PAH$^+$, PAH$^0$ and  VSG.
Since the 7.7/11.3 $\mu$m ratio is one of the main peculiarities of the observed spectra
and the problem affects all our fits, it may indicate
that the adopted templates,  in particular the  PAH$^0$ component, are not fully adequate.

\begin{table*}
\centering
\caption{Relative contribution in $\%$ of the different dust features components to the
integrated fluxes of the ETGs.}
\begin{tabular}{l|cccccccccc}
\hline\hline
\\
Galaxy& PAH$^0$&PAH$^+$&8.2BF$^a$&12BF$^a$&BF$^a$&VSG&L\\
&&&&&(8.2+12)&&PAH$^{0}$\\
\hline
NGC~1297&61&-&9 (11)&26 (30)&35 (41)&-&4\\
NGC~5044&43&5&10 (11)&26 (34)&36 (45)&9&7\\
NGC~6868&35&-&12 (13)&36 (38)&48 (51)&7&10\\
NGC~7079&58&-&8 (10)&24 (29)&32 (39)&-&10\\
\hline
\end{tabular}

Notes:  $^a$ Values in parenthesis correspond to the relative contribution in $\%$ of
the broad features components to the integrated fluxes of the ETGs in the
case of the Virgo stellar continuum template subtraction, see text for details.
\label{table5}
\end{table*}

\begin{table*}
\centering
\caption{Observed lines ratios, UV magnitudes and derived quantities}
\begin{tabular}{l|ccccc}
\hline\hline
&{N1297}&{N5044}&{N6868}&{N7079}\\
\hline

$[\rm{SIII}]18.7/[\rm{SIII}]33.5$&$0.37\pm0.10$&$0.35\pm0.15$&$0.36\pm0.10$&$0.38\pm0.12$\\
H$_2$S(3)/H$_2$S(1)&$0.66\pm0.09$&$1.49\pm0.30$&$2.85\pm0.21$&$1.32\pm0.15$\\
$[\rm{NeIII}]15.5/[\rm{NeII}]12.8$&0.61$\pm 0.02$&0.52$\pm 0.01$&0.74$\pm 0.01$&0.91$\pm 0.02$\\
$[\rm{SIII}]33.5/[\rm{SiII}]34.8$&$0.22\pm0.02$&$0.21\pm0.01$&$0.80\pm0.04$&$0.98\pm0.07$\\
\hline
$m_{1516{\AA}}$&-&20.05$\pm$0.12&19.97$\pm$0.10&21.10$\pm$0.14  \\
\hline
$n_{e}$($\rm{cm}^{-3}$)&0.1 -- 400&0.1 -- 400&0.01 -- 400&0.1 -- 400\\
G$_{\rm{FUV}}$(Habing units)$^{(a)}$&--&0.51&0.54&0.20\\
$\rm{G_{FUV}T_{gas}^{1/2}/n_e}$($K^{1/2}\rm{cm}^3$)&--&$<100$&$<110$&$<50$\\
\hline
\end{tabular}

Notes:  $^a$ The flux of the Habing fields, $G_{FUV}=1$, equal to $2.3 \times 10^{-3}$ erg cm$^{-3}$ s$^{-1}$ at $\lambda \sim 1530$ {\AA} \citep{mathis83}
\label{table6}
\end{table*}





Tielens (2008) points out that the PAH$^0$ template
 shows more emission in the 6 -- 9 $\mu$m region than that expected from theoretical calculations and laboratory
 studies of neutral PAHs. The same author suggests two different possibilities to explain this discrepancy: (a) the
 PAH$^0$ template
could represent the particular mixture of PAHs, which could be
 dominated by neutral PAHs
but with some, likely small, contribution from ionized PAHs,  and/or (b) it could also reflect the effect of the size
on the intrinsic spectrum of neutral PAHs. The latter point might  indicate that the emitting
PAH mixture in our ETGs has a size distribution more biased toward large and neutral PAHs,
than that one characterizing the PAH$^0$ component.

To bypass these difficulties we also include in our set of templates
 the  theoretical spectra  of a mixture of ten large, compact, and highly symmetric neutral, cationic and anionic,
PAHs, L-PAH$^0$, L-PAH$^+$, L-PAH$^-$, respectively, 
derived by \citet[][see their Figure 6]{bau08}. In order to compare those theoretical \emph{absorption} PAH spectra with observed PAH \emph{emission} spectra, band shapes, widths and shifts  inherent to  the emission processes must be taken into account \citep{bau08,bau09}. Therefore, we followed the recommendations of \citet{bau08,bau09} concerning the fitting of astronomical PAH spectra, and we red-shifted  the computed spectra by a value of 15 cm$^{-1}$, and used a bandwidth of 30 cm$^{-1}$ for the bands short-ward of 9 $\mu$m, and 10 cm$^{-1}$ for the bands long-ward of 10 $\mu$m.

The results of the new fits are shown in Figure \ref{fig9}, where the
 11.3$\mu$m feature is now well-fitted in all cases. In Table \ref{table5},
 we report the corresponding fractional contributions
of the single PAH templates to the best fits. The PAH spectra of our ETGs are well reproduced with only a few dust feature components: the PAH$^0$ and L-PAH$^0$, the broad feature components (i.e. 8.2BF + 12 BF) and the VSGs. Particulary stricking is the evident
deficit of the PAH$^+$ emission component, which is the main contributor in HII regions
and normal galaxies, see Figure \ref{fig10} (e.g. DL07, Tielens 2008 and references therein).
 Even if  the
PAH$^0$ and the L-PAH$^0$ components dominate the total PAH emission, suggesting that the main PAH contributors could be likely the neutral PAHs, we cannot completely discard the presence of ionized PAHs in the PAH mixture. That is because, on the one hand, the PAH$^0$ component can contain some contribution from ionized PAHs (Tielens 2008), and on the other, we are using a L-PAH theoretical spectra which was computed without including  the temperature cascade effects. \citet{boe10} show  that the inclusion of the PAH temperature cascade in the  calculations affects the relative band intensities.  In that case, the intensities of the C--H and C--C--C bands increase significatively with respect to the C--C bands (see Figure 18 in Boersma et al. 2010), and therefore the fractional contribution of large ionized PAHs to the PAH mixture in our galaxies could be greater than expected from the previous analysis. However, even when some contribution from ionized PAHs could be expected, the main results of our analysis still remains, i.e. the PAH spectra of our ETGs are dominated by emission from large  and neutral PAH molecules. Moreover, \citet{kaneda07a} reported unusual 7.7/11.3 $\mu$m inter--band ratio in addition to no-significant PAH emission at 3.3 $\mu$m in the near to mid-IR spectrum of the ETG NGC~1316. As the 3.3 $\mu$m emission is usually attributed to emission from small, neutral PAHs, Kaneda et al. conclude that the overall near to mid-IR PAH spectrum can be explained by emission from large and neutral PAH molecules.


\subsection{The Broad Feature components}

The other important contribution that we find in our galaxies is the broad  feature (BF) components (i.e. 8.2BF + 12BF) which is, in all cases, higher than  30 per cent, and even
higher than the neutral component in NGC~6868.

Those components are needed to fill the broad 7 -- 9 $\mu$m and 11 -- 12 $\mu$m features. In particular, the 8.2BF component would fill out the emission \textit{excess} in the 6 -- 9 $\mu$m spectral region mentioned in Section 3.2.1.
At first glance, it could seem that by subtracting the passive ETG template of,
we have artificially enhanced the broad 7--9 $\mu$m feature, and that, as a consequence,
 we need to explain its strength by invoking a {\it broad} emission feature, identified as
the 8.2BF component.
First of all, our objects are ETGs and a large fraction of their stellar populations
must be  old and quite metal rich (see Table \ref{tableape} in the Appendix), thus the presence of the
old stellar template is well justified.  In addition we adopted a template of the
passive galaxies in the field, which have a less pronounced 8$\mu$m dip than Virgo
cluster elliptical galaxies (see Figure \ref{fig1}).
We  also checked  how the selection of a particular  template
of an old stellar population influenced our results.
We repeat the analysis by using a template based on passive ETG members of the
 Virgo cluster. The results of the analysis are indicated  in parenthesis in Table \ref{table5}.
Finally, the BF component has an even larger broad contribution
around 12$\mu$m (12BF component) and, although in this region the fit is not perfect,
the presence of this component strongly reduces the discrepancy with the observed spectra.
 As a further check
we analyzed with the same technique the NGC~1482 spectrum
 (see Figure \ref{fig10}) which, at odds with our ETGs,  is well reproduced by a mixture
dominated by PAH cations, as expected in an HII galaxy (eg. DL07), while the
BF component is very small. In particular, the 8.2BF component is negligible. We then suggest that the presence of the BF component is not an artifact of the analysis.

So far, the BF features have been detected only in a few objects, likely C-stars in transition from
the  AGB  phase to the PN phase \citep[e.g.][]{buss93,peeters02,sloan07,joblin08}. However, \citet{sloan05} report broad and prominent features peaking at $\sim$ 8.3 $\mu$m in the MIR spectra of
 some proto-planetary disks around T-Tauri stars, and suggest that those broad features are the same
as those found in the MIR  spectrum of evolved stars (C-class objects from Peeters et al. 2002). \citet{bouwman08}
 and \citet{berne09} instead suggest that the broad features in T-Tauri stars present different peaks and widths than those of the C-class objects. Bern\'e et al. (2009) argue that, since the chemical conditions in proto-planetary disk and evolved stars are different, there is no reason to suppose that the species causing the broad features in proto-planetary disks are the same as those producing the broad features in evolved stars. Since the spectra of our galaxies are well fitted with the parameters of 8.2BF and 12BF templates, built from the spectrum of  the C-class object prototype, we assume that the BF emission in our galaxies is caused by the same material present in the C-class objects.

The  actual carriers of the BF emission are still unknown. These features are commonly attributed to carriers rich in aliphatic material \citep[e.g.][]{goto03,sloan07,joblin08}. However, Cami and coworkers find that the MIR spectrum of the C-class  prototype  object, IRAS 13416-6243 can be well reproduced with a mixture of small PAH spectra from the NASA Ames PAH spectral database (Cami et al. in preparation, see Figure 11 in Tielens 2008).
Regardless of the  actual carrier, which is out of the scope of this work, the key point here is that this material seems to be just created in the circumstellar envelopes of C-rich stars  and later processed and/or destroyed by the stronger ISM UV fields. This will be later discussed  deeper in Section 6.3.
In any case, our tentative detections of the aliphatic bands at 6.85 $\mu$m and at 7.25 $\mu$m  \citep[e.g.][]{chiar00,sloan07}
 in the spectra of NGC~5044 and NGC~6868 (see Figure \ref{fig10})  favour the aliphatic material
 interpretation. In that scenario, the BF spectra should arise from mixtures of aromatic and aliphatic
 carbon that have not yet been processed by intense ultraviolet radiation,
allowing the more fragile aliphatic bonds to survive \citep[e.g.][]{goto03,sloan07,joblin08,tielens08}.
A possible  evolutionary scenario for such features is also proposed: in cool and less evolved
 C-stars (in the post-AGB phase), the emission is dominated by this material. In a
hotter ISM environment the feature disappears because
the aliphatic material has  been processed into aromatic material \citep[e.g.][]{goto03,sloan07,joblin08}.
 In this way,  the presence of the BF features in our spectra would indicate the presence of pristine material.

\section{The interstellar radiation field}

The analysis performed above suggests that the unusual PAH inter-band ratios seen in ETGs are due to changes in the size and charge distributions of the PAH  mixture, which would be composed mainly of  large and neutral PAH molecules. In this section, we consider the  physical conditions in the ISM and whether or not they can account for such a PAH mixture.

\subsection{The PAH charge distribution}
The charge state of PAHs is mainly determined by the ionization parameter, G$_{\rm{FUV}}\rm{T_{gas}^{1/2}}/\rm{n_e}$ \citep[see][]{bakes01,tielens05} where $\rm{G_{FUV}}$ is the integrated far ultraviolet (900 -- 2000 {\AA}) radiation field in
Habing radiation field units, $\rm{T_{gas}}$ is the gas temperature in Kelvins and $\rm{n_e}$ is the electron density in $\rm{cm^{-3}}$. In order to estimate the values of the ionization parameter for our objects, we use the values of the [SIII]18.7/[SIII]33.5, the $H_2S(3)/H_2S(1)$ line ratios and the \textit{GALEX} far ultraviolet ($\lambda =1516$ {\AA}, $\Delta\lambda =268$ {\AA}) magnitudes within the central 5 arcsec from \citet{marino10}, all reported in the Table \ref{table6}, to  quantify the electronic density, the gas temperature and the integrated far ultraviolet  radiations field, respectively. The derived [SIII]18.7/[SIII]33.5 line ratios are compatible with  mean interstellar electronic densities, $\rm{n_e}$ between 0.1 and 300 cm$^{-3}$ \citep{dale06}. \citet{turner77} showed that $H_2S(3)/H_2S(1)$ line ratios ranging from 0.6 to 3 correspond to gas temperatures of 150 -- 400 K. Finally, by assuming the distances reported in Table \ref{tableape}, the FUV  magnitudes correspond to  average intensities of the interstellar FUV radiation field, in Habing units, of $\rm{G_{FUV} = 0.51, 0.54, 0.20}$ for NGC~5044, NGC~6868, and NGC~7079, respectively, which are about half the Galactic value. In the last row of Table \ref{table6}, we also display the corresponding  ionization parameters, G$_{\rm{FUV}}\rm{T_{gas}^{1/2}}/\rm{n_e}$, calculated for those three galaxies. The derived values are in  agreement with the expected values from  the quantification of the 6.2/11.3 $\mu$m ratio and the 7.7/11.3 $\mu$m ratio as a function of the ionization parameter given by \citet{flagey06} and/or by \citet{galliano08}.

We roughly estimate the fraction of neutral PAHs by following  \citet{tielens05}. In this case, the values of the averaged ionization parameter  of our objects (i.e. G$_{\rm{FUV}}\rm{T_{gas}^{1/2}}/\rm{n_e}\sim 100$ ) are compatible with a PAH mixture where even the 99\% of the PAHs as large as $\sim 1000 $ C-atoms are neutral. Lower values of the ionization parameter would indicate that even larger PAHs would be neutral.

\subsection{The PAH size distribution}
\citet{Panuzzo10} show that a diagram of the $[\rm{NeIII}]15.5/[\rm{NeII}]12.8$  vs. the
$[\rm{SIII}]33.5/[\rm{SiII}]34.8$ ratio allows the different excitation mechanisms in galaxies, i.e. starburst
photo-ionization, AGN photo-ionization, and shock heating, to be distinguished. Using this diagram,  the values of those line ratios for our ETGs, listed also in Table \ref{table6}, are accounted for by the shock+precursor models from \citet{allen08}, with
shock velocities $<150$ km s$^{-1}$, pre-shock gas densities
between 0.01--500 cm$^{-3}$ and solar abundances, indicating that relatively fast shocks in a low density medium is the primary powering mechanism in those LINERs. \citet{micelota2} analyze PAH processing in shocks by collisions with ions and/or electrons, and show that  PAHs with $\rm{N_C < 200}$ are completely destroyed by shocks with velocities $\sim 100-150$ km s$^{-1}$, and only PAHs with sizes  $\rm{N_C > 200}$  could partially survive for a time of $\lesssim 1-4\times10^8$ yr. 
 Fast shocks could be originated from jet-driven flows or accretion onto a massive black-hole, \citep[e.g.][]{nagar05,dopita97}, or from turbulent motion of gas clouds, either injected by stellar mass loss \citep[][]{bregman09}
or accreted, within the potential well of the galaxy.
However, independently of the origin, their presence and the low average UV radiation field in the post-shocked interstellar medium  could account for the changes in the size and charge distribution of the PAH mixture in our ETGs. The short lifetimes of the PAH molecules in such environments  and the fact that we see PAH features in the MIR spectra, require either mechanisms to protect the PAHs from destruction by shocks or a continuous supply of carbonaceous material into the ISM.

\section{Discussion}
\label{sec:discusion}
Although reminiscent of star forming galaxies,
the PAH spectra seen in our ETGs show remarkable differences
to such objects (see items in Section 3.2).  In addition, PAH molecules are expected to be  destroyed in a few $10^{6-7}$ years in the harsh environment of the ETGs. Hence, the presence of the PAHs in the elliptical galaxies seems to be incompatible with their passive interstellar environments dominated by hot plasma and/or shocks \citep[e.g.][]{bregman92,micelota1,micelota2,Panuzzo10}. So, the key questions to answer are (1) why do we see  PAH emission in ETGs, i.e. what is the source of the PAHs in our objects, and (2) why do the PAH spectra present that unusual shape?.

There are several possibilities to explain this anomalous emission.
We consider them in the following sections.

\subsection{Unusual emission due to the presence of an AGN}
\citet{smith07} noticed that all galaxies of the SINGs sample with peculiar
7.7/11.3 $\mu$m inter--band ratios are LINERS, and we also verified that all these galaxies are ETGs.
They suggested that the peculiar ratios may be due to the
presence of low level AGN activity. In that case, the  hard UV radiation field
would preferentially destroy small PAH molecules,
emitting at shorter wavelengths, and leave unaffected large PAHs,
which are thought to emit mainly at $\lambda > 11 \rm \mu m$.
This effect could change the PAH size distribution even at large distances
from the center \citep{voit92}. Even though this explanation may be correct to some extent,
it cannot be the \emph{only} reason for the observed shape of the continuum subtracted spectra. A significant emission at 6--9 $\mu$m
could still arise if
large PAHs  exist in the ionized state \citep[e.g.][]{bau08,bau09,boe10}.
In addition, the source of PAH molecules is unknown in the Smith et al. scenario.

\subsection{Emitting material recently accreted}
Another possibility has been advanced by \citet{kaneda07b,kaneda08}.
They suggest
that the unusual 7.7/11.3 $\mu$m inter--band ratio is due to a
recent encounter with a gas-rich galaxy
that supplied the ISM with a dust mixture biased toward large, neutral PAHs.
A large fraction of neutral PAHs could survive the
sputtering effects of the hot ISM because they were originally formed
on the mantle of the dust grains.
Among their arguments advanced to support this possibility we recall:
a) the observed missing anti-correlation between equivalent width (EW) of PAHs and
the X-Ray to B-band luminosity ratio, L$_{\rm{X}}$/L$_{\rm{B}}$, implying a lack of relation between the  PAH sources
(stellar mass loss rates) and the PAH sinks (X-rays).
b) The observed missing correlation between the 6$\mu$m continuum emission and the PAH (11.3$\mu$m) emission,
implying a lack of direct relation between PAH emission and stellar populations
of the galaxy.
c) The correlation between PAH emission and continuum emission around 35$\mu$m,
implying a relation between PAH and dust emission.
Concerning the first two points, we notice that only carbon-rich stars (C-stars) can provide
carbonaceous material, and that these stars are typical of intermediate age populations.
Their presence in an old system like our ETGs would be related to a rejuvenation
episode and not to the entire stellar population. One would therefore not expect a correlation
between PAH emission and the 6$\mu$m continuum.
Furthermore, the EW of PAHs
do not depend only on the abundance of carriers but also on the intensity of the exciting radiation,
which is mainly UV in the case of the of neutral PAHs \citep[e.g.][]{tielens08}. This UV flux
can vary from galaxy to galaxy as evidenced by GALEX \citep[e.g.][]{buson09}.
Finally, we notice that point c) only indicates that  there is a common origin for the exciting flux
of PAHs and dust.

Another major difficulty faced by this model is that,
in many cases, there is no  evidence that the last accretion episode
was sufficiently recent to allow the carriers to survive destruction by sputtering
or shattering mechanisms. Thus, the authors
invoke a mechanism that preserves the PAHs in the mantle of dust grains.
Apart from these considerations, we notice that
it remains unexplained why
this mixture should be biased toward neutral PAHs, considering that the
\textit{accreted material} should be biased toward the ionized PAHs, as is the ISM in normal
star forming galaxies.

\subsection{Emitting material continuously supplied by a population of C-stars}

The BF component, up until now, has been observed mainly in
carbon-rich pre-planetary nebulae \citep[e.g.][]{peeters02,sloan07}.
The prototype  is IRAS 13416-6243, a carbon star in transition from
the AGB phase to the planetary nebula phase. This raises the possibility that a population of such stars is responsible of the BF features observed in ETGs. However,
the comparison with our isochrones shows that this cannot be the case
 because the
AGB-PN transition phase is very brief,
about two orders of magnitude less than that required to produce the observed features in the integrated spectrum.

Nevertheless, we advance the hypothesis that a population of C-stars is the source of freshly synthesized carbonaceous material.
In our scheme, amorphous carbon  created within the
circumstellar envelopes of C-rich stars \citep[e.g.][]{whittet92}, is continuously ejected into the ISM and then shocked  and exposed to a weak UV radiation field, as described in Section 5.  

Such a scenario explains in a more natural way the origin of the  emission in these ETGs.
In our case, carbonaceous material is \emph{continuously}
supplied by the mass-losing carbon stars, whose origin must be
a recent rejuvenation episode.
The timescale for the process to occur is provided by the age of the carbon star progenitors,
which can vary from a few 10$^8$ yrs to a few 10$^9$ yrs, the latter timescale
depending on the chemical composition.
We thus do not need to advocate a mechanism to preserve
these  molecules from destruction (for PAHs several 10$^7$ -- $10^8$yrs),
as  Kaneda et al. (2008) did,
to explain the delay between the actual observation
and the epoch of most recent merging or of the gas capture episode.
In our model this delay is simply provided by the stellar evolution clock.

It is interesting to notice that such a population of carbon stars
should leave a signature in the MIR spectrum of the galaxy, due
to the emission of their carbon rich circumstellar envelopes \citep{bres98}
This emission is similar in shape and peak temperature to
the hottest residual continuous
component left after the subtraction of the stellar population template.
If this is the case, we can make a rough estimate of
the number of carbon stars responsible for this emission.
Assuming for such stars L=10$^4$ L$_\odot$  and that the light is almost
completely reprocessed by an envelope of optical depth $\tau_{1{\mu}m}\sim~10$,
 we obtain for NGC~5044 $\sim$ 15000 C stars.
This number would require a rejuvenation episode
involving $\sim$1 per cent of the sampled mass of the galaxy,
about 1 Gyr ago, which is a typical value found among field galaxies \citep{L00}.

In summary, our model can explain both
the low 7.7/11.3 $\mu$m ratio and
the presence of BF families as well as
relax the strong constraint imposed by the sputtering time on the epoch
of the gas accretion.

Thus we argue that the anomalous PAH spectrum
is a direct signature of the presence of an intermediate age
carbon star  population.

\section{Summary and conclusions}
\label{sec:conclusions}

We have analyzed the MIR spectra of four ETGs characterized by
an unusual PAH emission spectrum. These  galaxies, which belong to the sample
that will be fully presented in \citet{Panuzzo10}, show prominent PAH
complexes at 11.3 $\mu$m and at 17 $\mu$m, with abnormally low
7.7/11.3 $\mu$m inter--band ratios.

In contrast to star forming galaxies,
the IRS spectra of our ETGs are
dominated by stellar emission  at short wavelengths.
Thus, in order to investigate their continuum subtracted spectra,
we have first subtracted {\sl realistic templates}
of old stellar populations, derived by high signal-to-noise \emph{Spitzer} observations \citep{bressan06}.
After passive stellar template subtraction,
we have performed a detailed  analysis of the individual dust emission
components.

A surprising result is that the 7 -- 9 $\mu$m spectral region requires
features which are different from those found in normal late-type galaxies,
such as those discussed by DL07.
 In this spectral region some important
features look significantly wider and/or shifted toward longer wavelengths.

With the aim
of understanding all the above peculiarities
we have analyzed each spectrum by means of  dust feature templates \citep[see e.g.][]{joblin08} and  theoretical PAH spectra
\citep[see e.g.][]{bau08}. This
method has proved to be very efficient for the understanding of
the nature and origin of the PAH emission in Galactic objects
,but, to our knowledge, has never been used for external galaxies.

The result of this analysis shows that the ETG spectra are
dominated by two main components:
(1) the large and neutral PAH component, responsible for the peculiar 7.7/11.3 $\mu$m and 6.2/17 $\mu$m ratios,
and (2) the so called BF component, which produces  the broadening and the shift of the features in the 6--9 $\mu$m spectral range.
The PAH cations, that are the dominant family in
normal galaxies, are almost completely absent in these spectra, indicating that the PAH emitting mixtures in our galaxies have \textit{anomalous} size and charge distributions, biased to large and neutral PAHs.
In our sample, the BF components contribute approximately 30--50 per cent
of the total PAH flux between 6 and 14 $\mu$m and are responsible
for the observed shift and widening of the fitted Drude profiles
with respect to DL07.
To test whether this result is an artifact of the adopted procedure,
we have also analyzed the SINGs HII galaxy NGC 1482 \citep{smith07}.
In this case, the MIR spectrum
is well reproduced by a mixture
dominated by cationized PAHs, as is expected in an HII galaxy, while the 8.2BF contribution is negligible.

Up to now, the BF features have been observed mainly in
evolved carbon stars and have been associated with pristine
carbonaceous material \citep[e.g.][]{hony01,peeters02,berne07,sloan07,joblin08}. Thus,
our analysis provides convincing evidence that
we are seeing pristine carbonaceous components, still relatively
unprocessed by the ISM environment.
The most natural explanation we can advance is that
this material originates from a population of carbon stars.

We stress that in  the above analysis, abandoning the usual black body
template for the underlying old stellar populations, in favour
of a more realistic one derived from passive ETG galaxies, was a  fundamental step.

In order to account for the anomalous PAH mixture,  we characterized  the interstellar radiation fields in our objects by using mid-IR line ratios sensitive to the electron density, gas temperature and excitation mechanism. The results of this analysis suggest that shock models, with velocities between 100 -- 150 km s$^{-1}$, in a low density ISM, can account for the line ratios observed in our galaxies. If such shocks are present in our objects, they would first destroy the smaller PAHs on timescales of a few 10$^8$ yrs, and would help to maintain the observed size distribution. The extremely low values of the ionization parameter estimated for our objects lead to a predominantly neutral charge distribution.

Present data do not allow us to identify the actual impact of shocks on the dust size distribution, but the peculiar properties of the PAH emission in ETGs may be the result of an environment where the PAH molecules are processed in shocks, and excited by the weak UV radiation field of the old stellar population.



We propose that the unusual PAH spectrum
arises from the combination of two effects.
First, a rejuvenation episode that recently ceased
gives rise to a population of carbon stars that is now \emph{continuously}
feeding the ISM with fresh carbonaceous material.
From the comparison of  the residual warm dust component seen in the galaxies
with the circumstellar envelope dust emission of a typical carbon star,
we estimate that the rejuvenation fraction is about 1 per cent of the sampled mass.
We notice that kinematical studies of NGC~5044,
NGC~6868, and NGC~7079 show star/gas counter-rotation; a strong
indication of a past accretion event (see appendix).
Second, shocks process  the ejecta of these carbon stars, destroying
smaller PAHs. The weak UV radiation field of an old stellar population
excites the processed material, maintains a suitable ratio between
neutral and ionized PAHs and results in the {\it anomalous} spectral features.

It remains to be clarified whether the
presence of these shocks and the occurrence of
the rejuvenation episode are causally linked.
While there is a general  consensus that
AGN and star formation activity are
linked  \citep[see e.g.][]{bressan02,Bressan06b},
the situation is much less clear for LINERS.
However, if our interpretation is correct, our results
indicate that LINERS could also have hosted star formation
activity in the {\it recent} past, suggesting a
LINER/post-starburst connection.

\section*{Acknowledgments}

We are grateful to the
anonymous referee, whose comments and suggestions were very useful for improving the manuscript.
 We acknowledge partial financial support of the Agenzia Spaziale Italiana
under contract ASI-INAF I/016/07/0.
OV acknowledges the hospitality
 of the INAF-Osservatorio Astronomico di Padova and the financial
 support of the Mexican Conacyt project 49942-F. This work is
based on observations made with the \emph{Spitzer} Space Telescope
which is operated by the Jet Propulsion Laboratory,  California
Institute of Technology, under NASA contract 1407. .

\appendix

\section{Relevant properties of the galaxies so far}

The four galaxies considered in this paper belong to the
sample of 40 ETGs in low density environments investigated in
the mid-infrared by Panuzzo et al. (2010). This sample is
biased towards ETGs having emission lines in their optical spectra
\citep[see the analysis produced by][]
{Rampazzo05,Annibali06,Annibali07,Annibali10}.
It is known, however, that the presence of ionized gas in
ETGs is quite common both in the field as well as in clusters
\citep[see e.g.][]{Mac96,Sarzi05}.
Most of these ETGs  are classified as LINERs using optical diagnostic
diagrams (see Annibali et al. 2010).

In the Lick-IDS line-strength indices analysis,
performed by \citet{Annibali07}, the four ETGs discussed
in this paper were assigned quite different ages.

NGC~1297 is the oldest with a luminosity weighted age
of 15.5$\pm$1.2 Gyr, but only scant information exists
in the literature about this galaxy.

\citet{Annibali07} attributed to NGC~5044 the very uncertain
age of 14.2$\pm$10 Gyr. The galaxy, located in a rich group of galaxies
\citep{Tu88}, is rich in dust in the central 10\arcsec\ with a clumpy
distribution.  The ionized gas has a filamentary structure with an
extension of about 40\arcsec\ \citep{Mac96}. The gas velocity profile is
irregular, with many humps and dips, while the inner (within 1/3 of the
effective radius) stellar velocity profile is counter-rotating with
respect to the outer regions \citep{CMP00}. This galaxy, then, is a very
peculiar object, being a possible
merger/accretion remnant. \citet{Rickes04} studied the ionized
gas component in NGC~5044 suggesting the presence of both
a non-thermal ionization source in the central region and
an additional ionization source (possibly hot post-AGB stars)
in the outer parts.

NGC~5044 has a set of infrared observations. It has been detected by
IRAS and by ISO \citep[][and reference therein]{Ferrari02}. The
galaxy belongs to the \citet{kaneda08} sample and they detected PAH
features and also H$_2$ rotational emission lines and ionized
species using \emph{Spitzer}. \citet{Temi07} discussed IRAC and
MIPS \emph{Spitzer} observations of NGC~5044. They report
interstellar emission at 8$\mu$m detected out to 5 kpc as well as
70$\mu$m emission from cold dust exceeding that
expected from stellar mass loss. In view of the short sputtering
lifetime for extended dust ($\approx$ 10$^7$ yr) they conclude that
the extended dust cannot result from a recent merger with a gas-rich
galaxy. They support the view that the complex and highly fragmented
dust clouds are highly transient and created by stellar mass-loss in
the central $\approx$1 kpc. The dust clouds are intermittently
disrupted and heated by energy released by accretion onto the
central black hole (AGN Feedback).

\citet{Annibali07} provide luminosity weighted ages for NGC~6868 and
NGC~7079 being 9.2$\pm$1.8 Gyr and 6.7$\pm1.1$ Gyr,
respectively.

The Fabry-Perot observations of NGC~6868 \citep{Pl98} show that the
line--of--sight velocity field of the ionized gas component has a
velocity amplitude of $\pm$ 150 km~s$^{-1}$. \citet{CMP00} show that
along the axes at P.A.=30$^\circ$ and 70$^\circ$ the gas and stars have
similar kinematical properties, but along P.A.=120$^\circ$ the gas
counter-rotates with respect to the stellar component. \citet{Z96}
noticed the presence of an additional inner gas component which they
suggested could be due to the superposition of two unresolved
counterrotating components, one dominating the inner region, the other
dominating the outer parts. Stars also show a kinematically--decoupled
counterrotating core. The stellar velocity dispersion decreases towards
the galaxy center. The above kinematics suggest that NGC~6868
could have had a recent accretion episode which could explain the
relatively low luminosity weighted age.

\citet{Bettoni97} found that in NGC~7079 the gas is rotating in a direction
opposite to that of the stars (gas counter-rotation).

We may conclude
that, according to the current interpretation of the peculiar gas vs.
stellar kinematics found in NGC~5044, NGC 6868 and NGC 7079, accretion/merging
events could have occurred in these galaxies.

Both NGC~6868 and NGC~7079 has been observed in CO.
\citet{Huchtmeier92} observed with the SEST NGC~6868
and provide a CO upper limit. Assuming this value we obtain an
upper limit to the cold H$_2$ mass of $2.0\times 10^8$ M$_\odot$
\citep[adopting a galactic X factor X$_{\rm{CO}}=4.6$ M$_\odot$ (K~ km~s$^{-1}$
pc$^2$)$^{-1}$, see][]{Solomon87}. For NGC~7079,
\citet{Bettoni01} estimated a mass of cold H$_2$ of
$8.82\times 10^7$ M$_\odot$ within the central 2 kpc radius.

In Table \ref{tableape} we collect relevant properties of the
galaxies. Column (1) gives the name of the galaxy; column (2) the
morphological type; column (3) the class of activity; columns (4),
(5) and (6) the luminosity weighted age and the metallicity
and $\alpha$-enhancement computed from optical spectra by
\citet{Annibali07}. Column (7) lists the molecular
masses obtained from CO observations and by using the galactic X-factor.
Column (8) displays the X-ray luminosity reported in
\citep{Osullivan01}. In column (9) we give the luminosity distance taken
from NED using H$_{\rm{0}}$=73 km~s$^{-1}$Mpc$^{-1}$.

\begin{table*}
\centering
\caption{General properties of the sample}
\begin{tabular}{lcccccccc}
\hline
 NGC     & Type       &  Activity  & Age        & Z & [$\alpha$/Fe]             & M$_{H2}^{c}$ & log L$_{\rm{X}}$      &   D   \\
         &            &  Class     & (Gyr)           &   &                           & (M$_\odot$)             & (erg~s$^{-1}$) & (Mpc) \\
\hline
1297     & SAB0 pec:  &LINER       &15.5$\pm$1.2&0.012$\pm$0.001   &  0.29$\pm$0.04 &          &                &19.8\\
5044     &E0          &LINER       &14.2$\pm$10.0&0.015$\pm$0.022  &  0.34$\pm$0.17 &          &  42.74         &42.8\\
6868     &E2          &LINER       & 9.2$\pm$1.8&0.033$\pm$0.006   &  0.19$\pm$0.03 &$<2.0\times10^8$ &41.23    &37.5\\
7079     &SB0         &LINER       &6.7$\pm$1.1&0.016$\pm$0.003    &  0.21$\pm$0.05 &$8.82\times10^7$ &         &34.2 \\
\hline
\end{tabular}

Notes: (Col. 2) Morphological types are derived from RC3; (col. 3) Activity Class from \citet{Annibali10};
col.s 4-6 from \citet{Annibali07}; (col. 7) H$_2$ content from \citet{Huchtmeier92} and \citet{Bettoni01}
for NGC 6868 and NGC 7079 respectively; (col. 8) X-ray Luminosity from \citet{Ogle07}. (Col. 9) Distances adopted are
derived from {\tt NED}.
\label{tableape}
\end{table*}

\end{document}